\documentclass[11pt, english]{article}
\usepackage[utf8]{inputenc}
\usepackage{geometry}
\usepackage{float}
\usepackage{caption}
\usepackage{amsmath}
\usepackage{amssymb}
\usepackage{braket}
\usepackage{chemformula}
\usepackage{graphicx}
\usepackage{setspace}
\usepackage{esint}
\usepackage{subscript}
\usepackage[lofdepth]{subfig}
\makeatletter

\usepackage{array}
\usepackage{babel}
\usepackage{authblk}
\usepackage[section]{placeins}
\usepackage{xcolor}
\usepackage{soul}
\usepackage{array,booktabs}
\usepackage{relsize}
\usepackage{hyperref}
\usepackage{ulem}
\usepackage{cite}
\usepackage{ctable}
\usepackage{longtable}
\let\sub\textsubscript

\DeclareUnicodeCharacter{2212}{-}
\DeclareUnicodeCharacter{2264}{$\leq$}
\DeclareUnicodeCharacter{0327}{$\c{a}$}
\newcommand{\beginsupplement}{%
        \setcounter{table}{0}
        \renewcommand{\thetable}{S\arabic{table}}%
        \setcounter{figure}{0}
        \renewcommand{\thefigure}{S\arabic{figure}}%
        \setcounter{section}{0}
        \renewcommand{\thesection}{S\arabic{section}}%
        \setcounter{equation}{0}
        \renewcommand{\theequation}{S\arabic{equation}}%
        \setcounter{footnote}{0}        
     }

\title{Mapping Thermoelectric Transport in a Multicomponent Alloy Space}
\author{Ramya Gurunathan, Suchismita Sarker, Christopher K.H. Borg, James Saal, Logan Ward, Apurva Mehta, G. Jeffrey Snyder}
\date{January 11, 2022}

\begin{document}

\maketitle

\begin{abstract}
% [GSJ: Mention 'High Entropy Alloys' and/or Entropy Stabilization, basically the buzzwords used especially in TE papers using that strategy. This will both add interest and make it search-able]
% High  entropy alloy systems with multiple principal elements are considered attractive for thermoelectric materials because the mass and strain fluctuations introduced into the lattice can effectively scatter phonons and suppress the lattice thermal conductivity. [
Interest in high entropy alloy thermoelectric materials is predicated on achieving ultralow lattice thermal conductivity $\kappa\sub{L}$ through large compositional disorder. However, here we show that for a given mechanism, such as mass contrast phonon scattering, $\kappa\sub{L}$ will be minimized along the binary alloy with highest mass contrast, such that adding an intermediate mass atom to increase atomic disorder can increase thermal conductivity. Only when each component adds an independent scattering mechanism (such as adding strain fluctuation to an existing mass fluctuation) is there a benefit. In addition, both charge carriers and heat-carrying phonons are known to experience scattering due to alloying effects, leading to a trade-off in thermoelectric performance. We apply analytic transport models, based on perturbation and effective medium theories, to predict how alloy scattering will affect the thermal and electronic transport across the full compositional range of several pseudo-ternary and pseudo-quaternary alloy systems. To do so, we demonstrate a multicomponent extension to both thermal and electronic binary alloy scattering models based on the virtual crystal approximation. Finally, we show that common functional forms used in computational thermodynamics can be applied to this problem to further generalize the scattering behavior that is modelled.

% The expression for which can be further generalized by applying computational thermodynamics techniques used to calculate excess Gibbs energy, namely the Redlich-Kister polynomial and Muggianu model. 

% In the case of thermal transport, a recent reformulation of the Klemens alloy model provides a straightforward route to compute the thermal conductivity of multicomponent alloys. 
% [GJS: somewhat repeated above. getting to the point earlier will better keep readers. maybe trim or slightly reword]We find that the thermal conductivity is often lowest along the binary system with the largest mass contrast such that adding additional alloying elements is not necessarily beneficial from the standpoint of alloy scattering. Instead, we outline multicomponent alloying strategies that we would expect to reduce the thermal conductivity, all of which require taking advantage of multiple, orthogonal scattering effects. 
\end{abstract}

\section{Introduction}
%Sentences to include in Introduction
% [GSJ: Mention 'High Entropy Alloys' and/or Entropy Stabilization, basically the buzzwords used especially in TE papers using that strategy. This will add interest ]
High entropy alloy systems with multiple principal elements are considered attractive for thermoelectric materials because the mass and strain fluctuations introduced into the lattice can effectively scatter phonons and suppress the lattice thermal conductivity. Additionally, the multicomponent alloy space comprises a largely uncharted compositional territory for thermoelectric materials. Several recent reviews discuss the potential benefits of these multinary alloys to thermoelectric performance given their potential for unique, entropy-enabled characteristics\cite{JosephPoon2019, Liu2017, Gao2018}. For example, in addition to the substantial lattice distortions mentioned previously, entropy-stabilization also favors high crystal symmetry, which can lead to high electronic band degeneracy, and regions of extended solubility, which can facilitate doping.

Some recent experimental investigations of multinary alloys have given way to highly alloyed phases which substantially out-perform end-member compounds. Androulakis \textit{et al.}\cite{Androulakis2007} demonstrated that the large lattice mismatch and strain effects in \ch{Pb_{1-x}Sn_xTe}-PbS alloys leads to phase separation which introduce phase boundaries with nanoscale separation that scatter phonons and lead to an over 70\% reduction in lattice thermal conductivity relative to PbTe. In other instances, large electronic bandstructure changes in the multinary alloy space can produce peaks in thermoelectric performance, resulting, for example, from the convergence of electronic bands or opening of a band gap\cite{Xia2000, Ortiz2019}.

% For example, heavily alloyed \ch{ZrCoSb} and {HfCoSb} can open a band gap in these otherwise semimetallic compounds, leading large improvements in the Seebeck coefficient\cite{Xia2000}.

However, both charge carriers and heat-carrying phonons are known to experience scattering due to alloying effects. Analytic alloy scattering models, which predict how transport properties will vary with alloy composition, have been established both for carrier mobility as well as lattice thermal conductivity ($\kappa\sub{L}$) on the basis of a virtual crystal approximation (VCA) perturbation theory\cite{Makowski1973, Harrison1978, Mehrotra2011,  Klemens1955, Klemens1960}. If the end-member transport properties are well-established, these simple, physics-based expressions have been shown to capture the alloy effects on thermal and charge transport in numerous materials systems. However, these methods have thus far been applied only to binary alloy systems, often using the Makowski and Glicksman\cite{Makowski1973} expression for alloy scattering of charge carriers and the Klemens model for alloy scattering of phonons\cite{Klemens1955, Gurunathan2020MH}. Here, we show a multicomponent extension to these VCA-based approaches, allowing us to predict both the carrier mobility and lattice thermal conductivity across the full compositional range of high dimensional alloys, combining 3 or more end-member compounds. For thermoelectric systems, the electronic transport coefficient $\sigma_{E0}$ best captures the trade-offs between electrical conductivity, Seebeck coefficient, and the electronic thermal conductivity in thermoelectric performance. So, we additionally focus on how alloy mobility effects modify the $\sigma_{E0}$ coefficient. 

The multicomponent alloy scattering strength for phonon and charge carrier transport parallels the functional form used to compute the excess Gibbs free energy ($^EG$) in solid solutions. The $^EG$ quantifies the deviation from an ideal solution model and relates closely to the enthalpy of mixing. Two common techniques, the Redlich-Kister polynomial and Muggianu geometric method, are used to calculate the $^EG$ for high-dimensional alloys in sub-regular solution model, where the interaction energies between components are permitted to vary with composition of the alloy\cite{Hillert1980, HillertBook}. In the context of scattering, these methods allow one to treat scattering potentials that vary along the compositional range of the alloy, which can occur when the lattice properties of the end-members differ considerably. By borrowing these thermodynamic expressions, one can extrapolate more complex scattering profiles to the high-dimensional alloy space.

% For carrier mobility, the work of Makowski and Glicksman\cite{Makowski1973} provides a straightforward expression for alloy effects in a binary system, however, an extension to higher-order systems has not been presented. We develop a multicomponent extension to the alloy mobility model, based on straightforward materials inputs, by applying computational thermodynamics techniques used to calculate excess Gibbs energy, namely the Redlich-Kister polynomial and Muggianu model. In these thermodynamic expressions, higher dimensional activity coefficients, describing the interactions between three of more alloying elements, can be expressed as a sum over binary, pairwise interaction terms. We show that within the coherent potential and virtual crystal approximations used in analytic alloy models, the same relationships exists, such that the excess resistivity in multinary alloys can be computed from the alloy scattering parameters fit to the binary systems.

% In the case of thermal transport, a recent reformulation of the Klemens alloy model\cite{Gurunathan2020MH} provides a straightforward route to compute the lattice thermal conductivity ($\kappa\sub{L}$) of multicomponent alloys. 

From our multicomponent lattice thermal conductivity model, we find that the $\kappa\sub{L}$ is lowest along the binary system with the largest mass contrast such that adding additional alloying elements is not necessarily beneficial from the standpoint of alloy scattering. Therefore, the only way to reduce the $\kappa\sub{L}$ when entering a multinary alloy space is  to rely on a introducing a scattering mechanism that acts orthogonally or independently of mass defect scattering such as dislocation strain or boundary scattering.

Our analysis approach is applied here to predict the thermoelectric performance across the full compositional range of several pseudo-ternary half-Heusler alloys: \ch{(Ti, Zr, Hf)NiSn}, \ch{(Ti, Zr, Hf)CoSb}, \ch{(V, Nb, Ta)FeSb}, and \ch{(V, Nb, Ta)CoSn}. We additionally compare our predictions to the experimental mapping of thermoelectric properties in the quarternary PbTe-PbSe-SnTe-SnSe system described in Ortiz \textit{et al.}\cite{Ortiz2019}. Beyond thermoelectric materials, these easily-implemented analytic techniques can be applied to a wider range of electronic and thermal alloys, for applications such as microelectronics, thermal barrier coatings, or energy storage\cite{Ma2021, Wang2021HEM}, whenever transport properties are being tailored through solid solution mixtures.

\section{Theoretical Background}\label{subsct:theory_background}

The efficiency of a thermoelectric device is related to the material properties through the figure-of-merit $zT = S^2\sigma/(\kappa\sub{e} + \kappa\sub{L})$, where $S$ is the Seebeck coefficient, $\sigma$ is the electrical conductivity, and $\kappa\sub{e}$ and $\kappa\sub{L}$ are the electron and lattice components of the thermal conductivity, respectively. Each constituent of the $zT$ equation depends independently on doping level, such that the overall $zT$ is typically sharply peaked at an optimal carrier concentration. Therefore, comparing $zT$ values across different datasets and studies is only suitable when the measured samples are confirmed to be optimally doped. In contrast, the material quality factor $B$ is a doping-independent parameter that is correlated to the peak $zT$ achievable in a material, if the carrier concentration is optimized. The quality factor $B$ has the form

\begin{equation}
    B = \left(\frac{k\sub{B}}{e}\right)^2 \frac{\sigma\sub{E0}}{\kappa\sub{L}}T,
\end{equation}

where $\sigma\sub{E0}$, the electronic transport coefficient, characterizes electronic performance of the materials and is closely related to the drift mobility weighted by the density-of-states effective mass, also referred to as the weighted mobility: $\mu\sub{W} = \mu_0(m^*_{\mathrm{S}} / m_e)^{3/2}$. In this study, we focus on $B$ as a metric for thermoelectric performance. The following sections will discuss alloy scattering models for the lattice thermal conductivity and the electronic transport coefficient, emphasizing the generalization to multiple alloying elements.

\subsection{Klemens Alloy Thermal Conductivity Model}\label{subsct:klemens_model}

The alloy scattering model used here for thermal conductivity is the Klemens/Callaway model, which derives a point defect scattering strength $\Gamma$ based on the atomic mass and atomic radius differences between defect and host atoms. The basis of this model is the virtual crystal approximation, in which the true, defective lattice is compared to a reference, periodic lattice with the configurationally averaged atomic mass ($\langle M \rangle_c$) and radius ($\langle R \rangle_c$) at each lattice site. The later work by Tamura \cite{Tamura1984} emphasized the importance of sublattice-specific models when working with a multiatomic lattice (i.e. multiple atoms in the primitive unit cell). For each sublattice, the atomic mass and radius variance should be calculated when considering the population of atoms that can occupy that sublattice, including the host and any substitutional atoms. We will first focus on quantifying the mass variance scattering, as the radius variance can be treated analogously. First, the average mass variance on the $n\textsuperscript{th}$ sublattice is defined as:

% The final scattering strength $\Gamma$ is then simply the related to the stoichiometric average of the 

% In a multiatomic lattice, or a Bravais lattice with a basis greater than 1, it is helpful to build up $\Gamma$ starting from individual sublattices, indexed here by $n$. The scattering strength for sublattice $n$ mainly depends on the average mass variance, defined as:

\begin{equation}\label{eqn:mass_var}
    \langle \Delta M^2_n \rangle_c = \sum_i(x_i(M_i - \langle M_n \rangle_c))^2 
\end{equation}

%write out definition of <M_n>_c too

where $i$ indexes all species that can occupy the sublattice, $M_i$ is the species mass, and $\langle M_n \rangle_c$ is the average mass on the $n\textsuperscript{th}$ sublattice. 

A final stoichiometric average over all sublattices (indicated by $\langle \rangle_n$) in the formula unit must then be performed to get an aggregate value for the scattering strength due to mass contrast, or $\Gamma\sub{M}$. Finally, this averaged mass variance parameter is normalized by the average atomic mass in the system.

\begin{equation}
    \Gamma\sub{M} = \langle \langle \Delta M^2 \rangle_c \rangle_n = \frac{\sum_n d_n \langle \Delta M^2_n \rangle_c}{\sum_n d_n}
\end{equation}

Here, $d_n$ is the degeneracy or stoichiometric ratio of each sublattice in the formula unit. For the example compound \ch{Ti$_2$FeSb}, the sublattices are the \ch{Ti}, \ch{Fe}, and \ch{Sb} sites with degeneracies of 2, 1, and 1, respectively.

The total phonon scattering strength $\Gamma\sub{ph}$ value is simply the sum of the mass and radius (or strain) scattering parameters. This yields the following expression for $\Gamma\sub{ph}$ for a generic multiatomic lattice, containing an arbitrary number of point defects on each sublattice. 

\begin{equation}\label{eqn:total_gamma}
    \Gamma\sub{ph} = \Gamma\sub{M} + \Gamma\sub{R} = \frac{\langle \langle \Delta M^2_n \rangle_c \rangle_n}{\langle M \rangle_c^2} + \epsilon \frac{\langle \langle \Delta R^2_n \rangle_c \rangle_n}{\langle R \rangle_c^2}
\end{equation}

The $\epsilon$ parameter is known as the strain scattering parameter, and its value should be determined by the anharmonic coefficient $\gamma$ (i.e. the Gr\"uneisen parameter) as well as elastic properties such as the Poisson ratio of the host lattice.

Although the basis of this model is in point defect perturbations, in several materials systems, this model has shown to be predictive beyond the \textit{dilute defect} limit as it has been used to describe the full compositional range of several binary alloy systems\cite{Gurunathan2019, Gurunathan2020MH}. Additionally, the expressions we show here naturally extend to higher order systems in which multiple alloying elements can substitute on various sublattices in the system, as the procedure for calculating the average mass variance remains the same.

The Klemens model then relates the scattering parameter $\Gamma$ to the ratio of the defective thermal conductivity to that of the reference pure thermal conductivity $\kappa_0$. The $\kappa_0$ varies between endpoint thermal conductivities using Vegard's law.

\begin{align}
\frac{\kappa\sub{L}}{\kappa_0} = \frac{\mathrm{tan}^{-1}u}{u}
\hspace{20 mm}  
u^2 = \frac{(6\pi^5V_0^2)^{1/3}}{2 k_B v_s}\kappa_0 \Gamma.
\label{eqn:kappa_atan}
\end{align}

\subsection[Electronic Transport Function and its Extension to Alloys]{Electronic Transport Function $\sigma_{E0}$ and its Extension to Alloys}\label{subsct:background_sigmae0}

The electronic transport function is correlated to the maximum achievable thermoelectric power factor ($S^2\sigma$) for an optimally-doped sample. The electronic transport function ($\sigma_{E0}$) can be fit directly to the Seebeck coefficient $S$ versus conductivity $\sigma$ relation (Jonker plot), using the following relationship\cite{Snyder2020}

\begin{equation}\label{eq:WeightedMobilityApproxCombined}
\sigma\sub{E0}= \sigma
\left[\frac{\exp\left[\frac{|S|}{k_\mathrm{B} / e}-2\right]}{1+\exp\left[-5( \frac{|S|}{k_\mathrm{B} / e}-1)\right]}+ \frac{\frac{3}{\pi^2}\frac{|S|}{k_\mathrm{B} / e}}{1+\exp\left[5( \frac{|S|}{k_\mathrm{B} / e}-1)\right]}\right].
\end{equation}

%Both approaches give similar results based on comparisons done so far: see Purple Family Excel sheet
It is advisable to fit a single $\sigma\sub{E0}$ coefficient to a series of $S$-$\sigma$ pairs, measured at various carrier concentrations. Otherwise, computing the $\sigma\sub{E0}$ from a the $S$-$\sigma$ pair of an optimally doped sample (maximized $S^2\sigma$) is also suitable.

The Bardeen-Shockley equation predicts the carrier mobility by applying deformation potential theory, which considers the effect of phonon scattering on charge carriers. By treating phonons as a source of dilatational strain, and calculating the shift in conduction band or valence band energies in response to this strain (neglecting higher order effects like the change in effective mass), the deformation potential $\Xi$ is calculated in units of eV\cite{Bardeen1950}. The expression for $\sigma\sub{E0}$ based on deformation potential theory is given as:

\begin{equation}\label{eqn:sigmae0_defpot}
    \sigma\sub{E0} = \frac{2\hbar E\sub{L} N\sub{v}e^2}{3\pi m\sub{I}^*\Xi^2},
\end{equation}

\noindent where $E\sub{L}$ is the longitudinal elastic constant (often the bulk modulus is used), $N\sub{v}$ is the valley degeneracy, and $m\sub{I}^*$ is the inertial effective mass. Depending on whether the n-type or p-type $\sigma\sub{E0}$ is desired, the band properties of either the conduction or valence band are applied.

%Add citation to Nordheim's rule papers and Makowski and Glicksman
In our implementation, we assume that the end-member compounds in the alloy systems are described by Equation \ref{eqn:sigmae0_defpot}. Alloy compositions will then have additional mobility effects due to alloy scattering of charge carriers. The alloying elements will be assumed to be isovalent substitutions and therefore uncharged. The scattering potential, in this case, will be determined solely from changes in the lattice potential at the defect site (i.e. pure potential scattering), meaning no significant spin disorder. 

Here we will invoke the virtual crystal approximation once more, by comparing the disordered lattice to a reference, periodic lattice with the averaged lattice properties of the alloy components. The coherent potential approximation (CPA) is also frequently applied to this problem, and involves self-consistently determining the self-energy of electronic states in an alloy to determine their scattering rates. In the limit of small perturbations (low defect concentration or low site energy change), the CPA also reduces down to the virtual crystal approximation\cite{ZimanDisorder}. It is common to evaluate the CPA correction to virtual crystal band energies, and if it is determined to be small, the virtual crystal approximation is appropriate to use for determining alloy scattering rates\cite{Murphy-Armando2006}.

Unlike the previous section on phonon--point-defect scattering (Section \ref{subsct:klemens_model}), in which the variance is atomic mass and radius defined the scattering strength, the lattice quantity defining the scattering strength in this case is $U$, the on-site potential in a tight-binding representation. The carrier lifetime associated with alloy scattering is related to the average variance in $U$, the density of site defects $n\sub{d}$, and the 3D electronic density of states $g(\epsilon)$\cite{Heng2022}:

\begin{align}
    \tau^{-1}_{\mathrm{alloy}} &= \frac{2\pi}{\hbar} n\sub{d} V_0 (\Delta U)^2 g(\epsilon)\nonumber \\
    &= \frac{2\sqrt{2}}{\pi\hbar^4}V_0 (\Delta U)^2 m^{*3/2}(k\sub{b}T)^{1/2}\epsilon^{1/2}
\end{align}

The practical definition of $U$ has been a source of discrepancy. The original work of Makowski and Glicksman\cite{Makowski1973} on III-IV zinc-blende compounds used the bandgap to define $U$, while subsequent work by Harrison and Hauser\cite{Harrison1978} defined $U$ as the electron affinity, or the conduction band edge position relative to the vacuum level. However, neither definition has shown a very robust correspondence to binary alloy data, and so, in practice, the value of $\Delta U$ between two components is fit to the mobility data of binary alloy systems. 

% We will first show the relaxation time due to alloy scattering strength, and then discuss the expression $\sigma\sub{E0}$, which combines both phonon and alloy scattering of charge carriers using Matthiessen's rule. 

% A rather simple derivation is required to arrive at the squared matrix element term of the carrier lifetime, starting with the average potential of the disordered lattice with two components: $X$ and $Y$:

% \begin{equation}
%     \overline{U} = \sum_i x_i U_i
% \end{equation}

% Analogous to the previous case of the mass defect (Section \ref{subsct:klemens_model}), we can write the real-space scattering potential at a defect site with a change in lattice potential $\Delta U$:

% \begin{equation}
%     V(\mathbf{r}) = V_0\Delta U \delta(\mathbf{r}).
% \end{equation}

% Following the procedure used previously, the alloy scattering rate includes the site fraction of point defects ($n\sub{d}$) and the 3D electronic density of states ($g(\epsilon)$):

% \begin{align}
%     \tau^{-1}_{\mathrm{alloy}} &= \frac{2\pi}{\hbar} n\sub{d} V_0 (\Delta U)^2 g(\epsilon)\nonumber \\
%     &= \frac{2\sqrt{2}}{\pi\hbar^4}V_0 (\Delta U)^2 m^{*3/2}(k\sub{b}T)^{1/2}\epsilon^{1/2}
% \end{align}

As in the mass difference case, when the factor of $(\Delta U)^2$ is generalized to multiple defects indexed by $i$ with site fraction $x_i$ and site potential $U_i$, it can be written as the average $U$ variance compared to the virtual crystal site potential $\langle U \rangle = \sum_i x_i U_i$. Note that here we will use angular brackets $\langle \rangle$ to signify configurational averages over the component site potentials. A rearrangement of the average $U$ variance puts it into a form particularly convenient for the charge carrier alloy scattering problem since $\Delta U_{ij} = U_i - U_j$ values between two components are nearly always fit to binary alloy data. The scattering parameter for the multicomponent alloy is written as a sum over ``binary" terms, involving just two components in the alloy.

\begin{align}\label{eqn:Delta_U2_manipulated}
\begin{split}
    \langle \Delta U^2 \rangle_c &= \langle (U - \langle U \rangle)^2 \rangle \\
    &= \langle U^2 \rangle - \langle 2 U \langle U \rangle\rangle + \langle U \rangle^2 \\
    &= \langle U^2 \rangle - 2\langle U \rangle^2 + \langle U \rangle^2 \\
    &= \langle U^2 \rangle - \langle U \rangle^2 \\
    &= \sum_{i, j\neq i}x_ix_j(\Delta U_{ij})^2
\end{split}
\end{align}

In Wang \textit{et al.}, the following expression for $\sigma\sub{E0}$ is derived by including the effects of deformation potential phonon scattering and alloy scattering, combined using Matthiessen's rule\cite{Wang2012}. The resulting expression is then a modification of the reference pure ${}^\mathrm{p}\sigma\sub{E0}$ value shown in Equation \ref{eqn:sigmae0_defpot}. Note that ${}^\mathrm{p}\sigma\sub{E0}$ represents the properties of the virtual crystal, and is defined as the Vegard's law interpolation between end-member values (similar to $\kappa_0$ in Equation \ref{eqn:kappa_atan}).  

\begin{equation}\label{eqn:sigmae0_alloy}
    \sigma\sub{E0} = {}^{\mathrm{p}}\sigma\sub{E0}/\left(1 + A\sum_{i, j\neq i} x_ix_j(\Delta U_{ij})^2\right),\,  \mathrm{where} \; A = \frac{3\pi^2E\sub{L}V\sub{at}}{\Xi^28k\sub{B}T}
\end{equation}

\noindent In the coefficient $A$, quantities such as the longitudinal elastic constant $E\sub{L}$ and average volume per atom $V\sub{at}$ also vary according to Vegard's law between end-member values. Finally, to relate this problem back to the previous case on point defect scattering of phonons, let us define the electronic alloy scattering strength $\Gamma\sub{el}$ as:

\begin{equation}\label{eqn:gamma_AS}
    \Gamma\sub{el} = \sum_{i, j\neq i} x_ix_j(\Delta U_{ij})^2
\end{equation}

In contrast to previous expressions presented by Makowski and Glicksman\cite{Makowski1973}, Harrison and Hauser\cite{Harrison1978}, and Mehrotra\cite{Mehrotra2011}, Equation \ref{eqn:sigmae0_alloy} is generalizable to multiple alloying elements. We additionally note that the final form of Equation \ref{eqn:Delta_U2_manipulated} is similar to those used for excess Gibbs energy calculations in a multicomponent solution.

In the following section, we present an extrapolation scheme from binary alloy data to higher order systems based on the Muggianu model. A benefit of this extrapolation scheme is that it preserves the relationship between $\sigma\sub{E0}$ and alloy composition shown in Equation \ref{eqn:sigmae0_alloy} even if $\Delta U_{ij}$, itself, is allowed to vary with composition. In thermodynamic terms, an activity coefficient with composition dependence represents a subregular solution model used first by Hardy to model certain binary metallic systems\cite{Hillert1980, Ganguly2001}. In scattering terms, $\Delta U_{ij}$ values may vary in the terminal regions of the alloy if there are significant differences in bandstructure and host lattice properties between endmembers. So, rather  than adopting a fixed $\Delta U_{ij}$ values over the entire alloy range, it maybe preferable to allow this alloy scattering potential to vary with alloy composition. The compositional variation in alloy scattering potential has only been rigorously investigated via first-principles for \ch{Si_{1-x}Ge_x} alloys, where changes in $\Delta U$ with respect to $x$, determined through DFT or tight-binding bandstructures, were fit to polynomial expressions. In this particular example, only a light dependence on composition was observed such that the assumption of a uniform alloy scattering potential provided an adequate fit of mobility data\cite{Mehrotra2011, Murphy-Armando2006}. In Section \ref{subsct:quat_system}, we discuss a quaternary reciprocal IV-VI semiconductor system in which the composition dependence is more pronounced.

\subsection{Redlich-Kister Polynomial and Muggianu Model for Excess Quantities}

The Redlich-Kister polynomial is a simple change of variables that allows for the analytic calculation of excess Gibbs free energy ($G^{\mathrm{E}}$) in a ternary, or higher order, alloy system. The method is especially useful in a subregular solution model, in which the interaction parameter ($A_{ij}$) between two components $i$ and $j$ is allowed to vary with composition. Here, we'll adapt the notation and convention of Hillert\cite{Hillert1980, HillertBook}. The standard regular solution model for $G^{\mathrm{E}}$ is shown below for constant binary interaction parameters:

\begin{equation}\label{eqn:reg_solution}
    G^{\mathrm{E}} = \sum_{i, j\neq i}x_ix_jA_{ij}
\end{equation}

\noindent As shown, higher order systems can continue to be written in as a sum of binary terms of the form $x_ix_jA_{ij}$ within the regular solution model. However, once $A_{ij}$ is permitted to vary with composition, this relationship no longer holds true, largely because the symmetry of the problem is not maintained $1-x_i \neq x_j$. To address this, in an example ternary composition ($x_1$, $x_2$, $x_3 = 1- x_1 - x_2$), we can define a new set of parameters using a simple change of variables:

\begin{align}
    v_{12} &= (1 + x_1 - x_2)/2\\
    v_{21} &= (1 + x_2 - x_1)/2.
\end{align}

\noindent In a binary system (where $x_1 + x_2 = 1$) these parameters reduce to simply $x_1$ and $x_2$, but they retain the relationship $v_1 + v_2 = 1$ even for higher order systems. The interaction term $A_{ij}$ can then be written as the following power series up to any arbitrary power $n$:

%Take another look at this equation
\begin{align}\label{eqn:RK-polynomial}
    A_{ij} &= \sum_{k=0}^n{{}^kC_{ij}v_{ij}^{n-k}v_{ji}^{k}}\\
     &= \sum_{k=0}^n{{}^kD_{ij}(x_i - x_j)^k}\nonumber
\end{align}

\noindent The two functional forms shown in Equation \ref{eqn:RK-polynomial} are both valid because $v_{ij}$ annd $v_{ji}$ only contain $x_i$ and $x_j$ in the combination $x_i - x_j$. Next, as before, the excess Gibbs energy can be described using Equation \ref{eqn:reg_solution}, using this more generalized form of $A_{ij}$: 

\begin{equation}\label{eqn:RK_excess}
    G^{\mathrm{E}} = \sum_{i, j\neq i}x_ix_j\sum_{k=0}^n{{}^kD_{ij}(x_i - x_j)^k}.
\end{equation}

% Equation \ref{eqn:RK_excess} provides a route forward to develop a higher order extension of Equation \ref{eqn:nordheim}. It is important to utilize the Redlich-Kister polynomial because the binary scattering coefficient ($CU^2$) is also composition-dependent and includes several combined quantities, which each vary with composition according to Vegard's law. 

% \begin{figure}
%     \centering
%     \includegraphics[width=0.5\textwidth]{Figures/muggianu_schem.pdf}
%     \caption{The geometric Muggianu model provides a method to determine a ternary excess quantity (multicolored center composition) from a weighted sum of corresponding binary excess quantities. Binary compositions are shown along binary systems as red, blue, and green scatter points.}
%     \label{fig:mugg_schem}
% \end{figure}

\begin{figure}
    \centering
    \includegraphics[width=\textwidth]{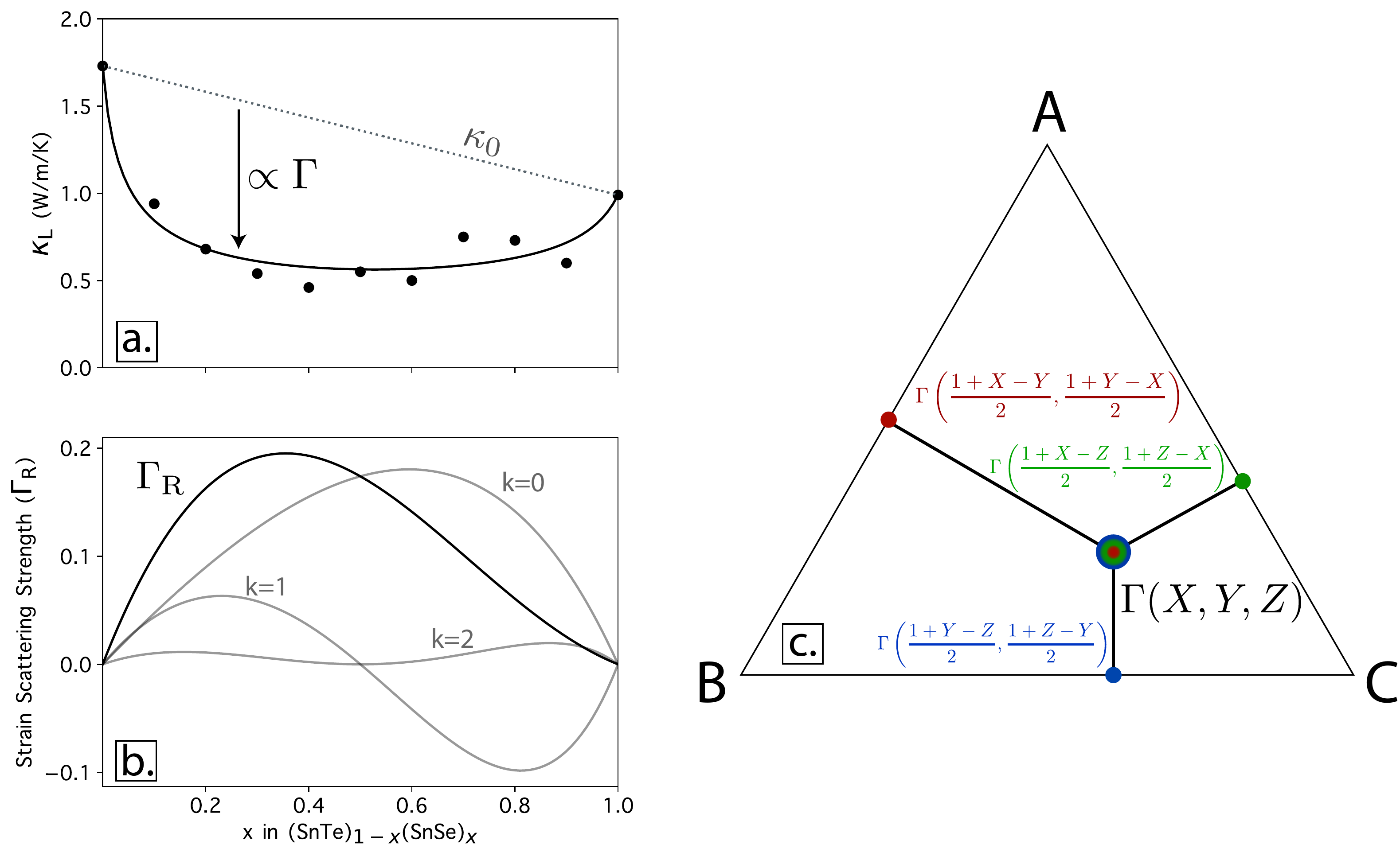}
    \caption{a) The scattering strength $\Gamma$ quantities the reduction in thermal (shown here) or electronic transport relative to a reference pure value that linearly interpolates between end-members. b) The Redlich-Kister polynomial can be used to model $\Gamma$ across the alloy region. Here, the first three terms of the polynomial ($k= 0,1,2$) are shown as well as the combined effect on $\Gamma$. c) The geometric Muggianu model provides a method to determine a ternary excess quantity (multicolored center composition) from a weighted sum of corresponding binary excess quantities. Binary compositions are shown along binary systems as red, blue, and green scatter points.}
    \label{fig:mugg_schem}
\end{figure}

To further simplify the model, we can apply the Muggianu method, referred to as a geometric model because it relies on the geometric construction illustrated in Figure \ref{fig:mugg_schem}. Here, ternary excess quantities are computed as a weighted sum of corresponding excess quantities along the binaries. Binary alloy systems are often well-studied experimentally and so it is possible to fit descriptive alloy scattering models described by Equation \ref{eqn:sigmae0_alloy}. The Muggianu method then provides the most straightforward extrapolation into the higher order space. The Muggianu method defines high-dimensional (h.d.) excess alloy properties as:

\begin{equation}
    G^{\mathrm{E}}\sub{h.d.} = \sum_{i, j\neq i}\frac{x_ix_j}{v_iv_j}G^{\mathrm{E}}(v_i, v_j)
\end{equation}

In thermodynamic models for $G^{\mathrm{E}}$, ternary interaction terms of the form $x_ix_jx_kA_{ijk}$, which involve three different components of the alloy, may also be used. In scattering problems, these terms would represent multiple scattering events present at clusters of defects. In our applications of the Redlich-Kister polynomial, these ternary interaction terms are neglected. Therefore, all terms in the scattering potential contain the site fractions and lattice properties of just two components in the alloy. 

The Muggianu method reproduces the Redlich-Kister polynomial terms up to any power\cite{Hillert1980, HillertBook}. Relating this expression back to the problem at hand, the working formula for the higher order extension of the electronic alloy scattering strength (Equation \ref{eqn:gamma_AS}) is:

\begin{equation}
    \Gamma\sub{el}(x_1, x_2, \ldots, x_n) = \sum_{i}^{n}\sum_{j \neq i} \frac{4x_ix_j}{(1 + x_i -x_j)(1 + x_j - x_i)} \Gamma_{\mathrm{el}}^{ij}(\frac{1 + x_i - x_j}{2}, \frac{1 + x_j - x_i}{2}).
\end{equation}

Since the point defect scattering strength for phonons $\Gamma\sub{ph}$ has the same form, we can apply the Muggianu method in the same fashion. If, perhaps, the strain scattering parameter $\epsilon$ (see Equation \ref{eqn:total_gamma}) describing the sensitivity of the force constants to strain varies over the alloy compositional range, the Muggianu method can be applied to perform the extrapolation to multicomponent alloys. In Section \ref{subsct:quat_system}, we will discuss a quaternary reciprocal IV-VI semiconductor system in which the Redlich-Kister polynomials and Muggianu extrapolation are required to adequately describe the experimental data.

\section{Results and Discussion}

\subsection{Pseudoternary Half-Heusler Alloys}

Half-Heusler compounds represent a large class of semiconducting, thermoelectric materials, which tend to be limited in their performance by relatively high $\kappa\sub{L}$ values. As a result, alloying strategies to impede phonon transport are heavily sought after. We have investigated several pseudoternary half-Heusler systems with alloying on a single sublattice, including: \ch{(Ti, Zr, Hf)NiSn}, \ch{(Ti, Zr, Hf)CoSb}, \ch{(V, Nb, Ta)CoSn}, and \ch{(V, Nb, Ta)FeSb} (typically doped with 20\% \ch{Ti}). For simplicity, we will also label the alloying site as $X$. The methods described in Section \ref{subsct:theory_background} were applied to map out the transport function $\sigma\sub{E0}$, lattice thermal conductivity $\kappa\sub{L}$, and thermoelectric quality factor $B$ over the full pseudoternary compositional space. These models require that the transport coefficients, elastic properties, and electronic bandstructure properties are well-defined for the end-member compounds, and literature values are used.

\begin{figure}
    \centering
    \includegraphics[width=\textwidth]{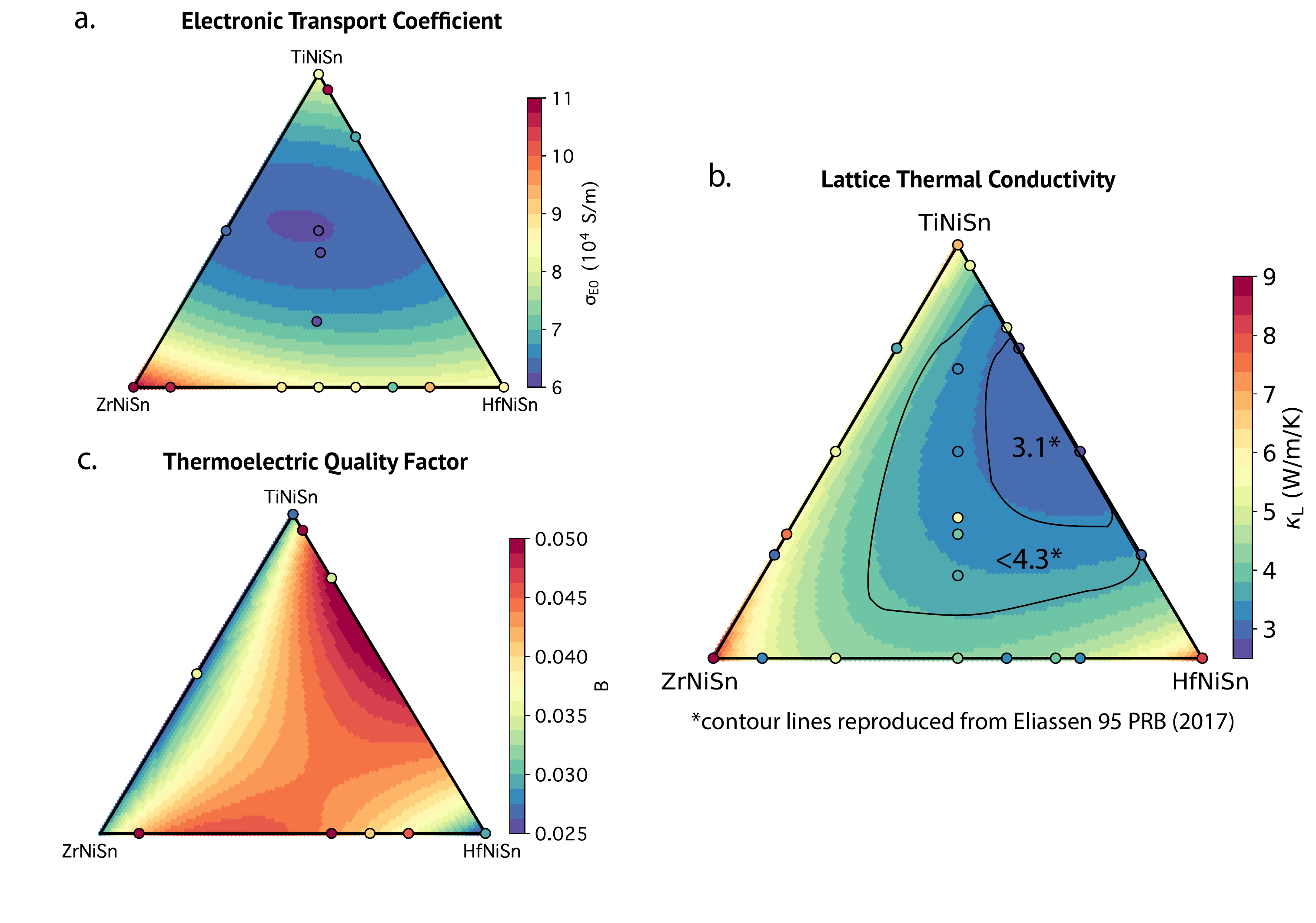}
    \caption{Alloy model predictions for the electronic transport function $\sigma\sub{E0}$, lattice thermal conductivity $\kappa\sub{L}$, and quality factor $B$ for the (Ti,Zr,Hf)NiSn system. Experimental scatter points are overlaid, and in cases where a composition was measured multiple times, the median value is plotted here. Contour lines from the DFT investigation by Eliassen \textit{et al.}\cite{Eliassen2017} are reproduced in panel (b) for comparison purposes. See Supplementary Section \ref{suppsct:hh_exp_data} for full data.}
    \label{fig:xnnisn_properties}
\end{figure}

Figure \ref{fig:xnnisn_properties} shows an example property mapping for the $X$NiSn system, and the remaining half-Heusler alloys are depicted in Supplementary Figure \ref{fig:hh_tern_grid}. Experimental $\kappa\sub{L}$ and $\sigma\sub{E0}$ scatter points obtained from the literature are overlaid for comparison and summarized in tabular format in Supplementary Section \ref{suppsct:hh_exp_data}. The half-Heusler often shows large variation in $\kappa\sub{L}$ and $\sigma\sub{E0}$ measurements, which may in part be due to variation in microstructure between samples. Of note, the alloy scattering model treats solid solution mixing, and so deviation between the model and experiment may indicate new scattering mechanisms stemming from microstructure or phase separation\cite{Schwall2013}. Still, in Figure \ref{fig:xnnisn_properties}, we plot the median value for compositions measured and reported multiple times in the literature, and the scatter points appear to follow the general trend predicted by the model. 

Inspection of the $\kappa\sub{L}$ maps for the four half-Heusler compound families shows that the lattice thermal conductivity is not, in fact,  minimized at the center of the ternary, where maximal compositional disorder would be expected. Instead, $\kappa\sub{L}$ is minimized along the pseudobinary with the greatest mass difference between components located along the right edge of each pseudoternary diagram shown. In fact, additional alloying with the intermediate mass component tends to reduce the overall point defect scattering.

These $\kappa\sub{L}$ mappings suggest that multicomponent alloying does not by necessity lead to suppressed thermal conductivity, a result corroborated by recent first-principles thermal conductivity calculations\cite{Yang2020, Schrade2017, Eliassen2017}. The \ch{(Ti, Zr, Hf)NiSn} $\kappa\sub{L}$ heatmap (Figure \ref{fig:xnnisn_properties}b) has been presented twice before from the standpoint of \textit{ab initio} density functional theory (DFT)\cite{Schrade2017, Eliassen2017}. Notably, these DFT studies continue to make the virtual crystal approximation when evaluating the alloy phonon bandstructures, using the compositional average of the atomic masses, harmonic, and anhamornic force constants. However, in contrast to the analytic expression here, the DFT-derived density-of-states, group velocities, anharmonic coefficients, and polarization vectors enter into the $\kappa\sub{L}$ calculation. We observe excellent correspondence between our model prediction and the DFT studies, which also show the minimum thermal conductivity along the (Ti,Hf) binary. Additionally, the computational work of Caro \textit{et al.} used non-equilibrium molecular dynamics and the Green-Kubo method to evaluate the thermal conductivity of Lennard-Jones alloys between fictitious elements \textit{A, B, C} and \textit{D}, which are assigned distinct atomic mass, radius, and cohesive energy\cite{Caro2015}. A benefit of this model is that it does not require any assumptions about the phonon bandstructure or the scattering mechanisms. The thermal conductivity mapping of the quaternary alloy still, however, shows the lowest $\kappa\sub{L}$ along the binary with highest mass contrast. In each of the half-Heusler alloy systems investigated, mass difference scattering alone appeared to adequately fit the experimental data, and the fitted value for the strain parameter $\epsilon$ was very close to 0. This low strain scattering effect is compatible with conventional understanding of formation rules for multicomponent alloys. Large atomic size differences lead to insufficiently negative enthalpy of formation values and poor miscibility\cite{Zhang2008}. 

Despite the similarity in motif for the lattice thermal conductivity heatmaps, the quality factor plots for the four pseudoternary families (Figures \ref{fig:xnnisn_properties} and \ref{fig:hh_tern_grid}) show different regions of high performance materials, arising from the trade-off between thermal and electronic transport coefficients. The magnitude of the quality factor predictions are similar between the four families. The $X$FeSb is typically doped with large quantities of \ch{Ti} and so a 20\% \ch{Ti} content is used in the model. This family exhibits quality factor $B$ values about a factor of 3$\times$ higher than the other systems stemming from both high $\sigma\sub{E0}$ and low thermal conductivity (partially due to the addition of \ch{Ti}). Quality factor values are particularly high along the (Nb,Ta)FeSb pseudobinary (Figure \ref{fig:hh_tern_grid}(a-c)).

\subsection{Quaternary Chalcogenide System: \ch{Pb(Sn)Te(Se)}}\label{subsct:quat_system}

We additionally apply this alloy model to the experimental mapping of thermoelectric transport properties performed by Ortiz \textit{et al.}\cite{Ortiz2019} of the quaternary p-type \ch{PbTe}-\ch{PbSe}-\ch{SnTe}-\ch{SnSe} system. Although several previous works have investigated the 6 pseudobinary systems involving all pairs of these IV-VI compounds, this study is unique in that it reports on the multicomponent alloy space. In our analysis, we exclude values in the vicinity of \ch{SnSe}, which were shown in the original work to be the \textit{Pnma} rather than the \textit{Fm$\overline{3}$m} phase\cite{Ortiz2019}.

To perform the lattice thermal conductivity analysis, we first fit the Klemens alloy scattering model (Equation \ref{eqn:kappa_atan}) to each of the pseudobinaries. The fit strain scattering parameters $\epsilon$ vary greatly, ranging from $\epsilon = 1$ for the SnTe-SnSe system to $\epsilon = 84$ for the PbTe-SnTe system. This implies that the anharmonic and elastic properties of the lattice vary too greatly over the compositional range for a single value of $\epsilon$ to be suitable. Therefore, we allow $\epsilon$ to vary with composition according to the Redlich-Kister polynomial up to the quadratic term $k = 2$ (see Equation \ref{eqn:RK-polynomial}). In order to adequately fit the experimental data, it was important to use the Muggianu model to extrapolate the $\Gamma\sub{R}$ from the pseudobinary to the pseudoquaternary space. 

\begin{figure}
    \centering
    \includegraphics[width=\textwidth]{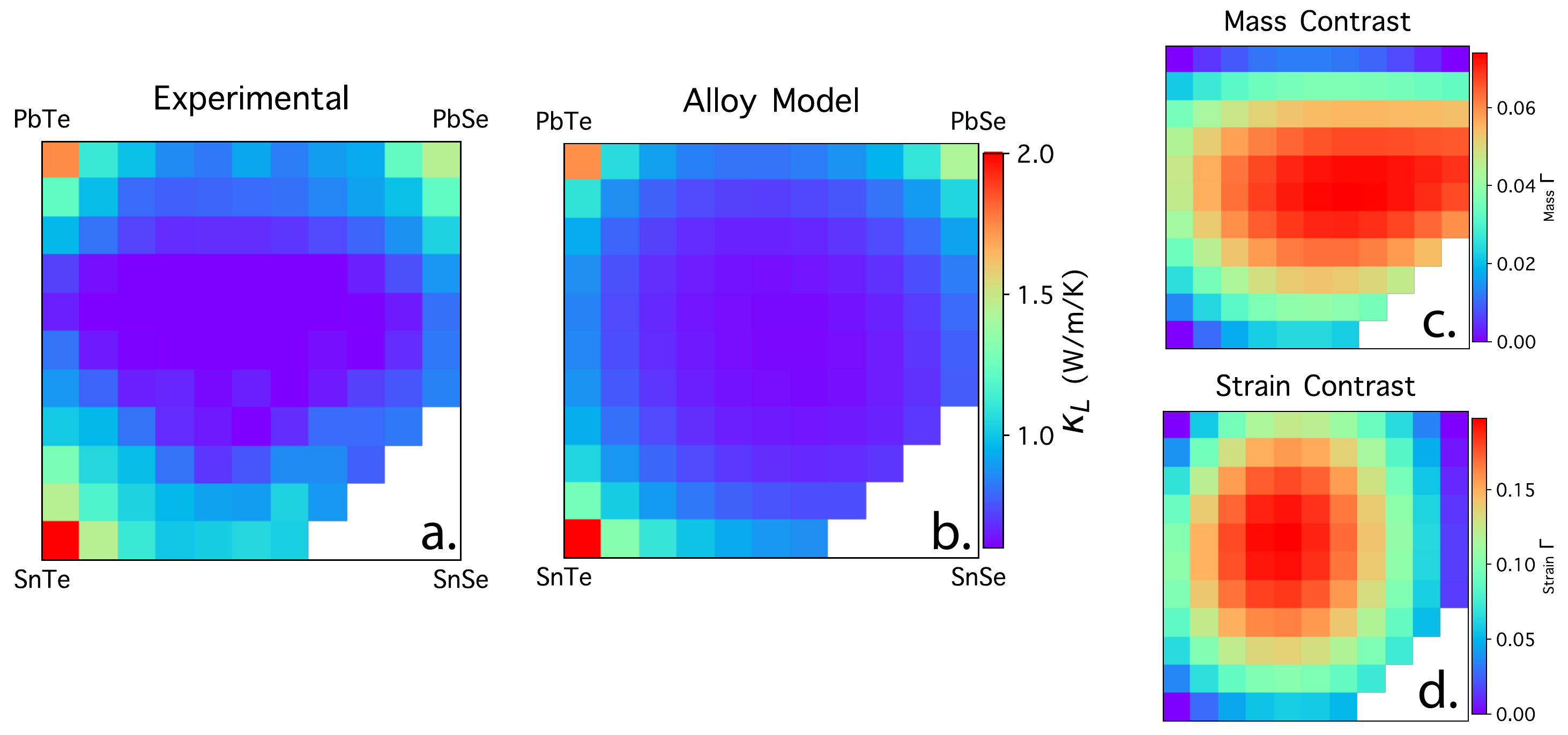}
    \caption{Lattice thermal conductivity heatmap in the quaternary IV-VI semiconductor system. Both the experimental measurements (a) and alloy model (b) show the thermal conductivity minimized near the equiatomic \ch{Pb_{0.5}Sn_{0.5}Te_{0.5}Se_{0.5}} composition. In this system, subsitution on both the cation and anion site yields a peak mass and strain contrast (c,d) near the center of the compositional space.}
    \label{fig:quat_chalc_kL}
\end{figure}

In doing so, we predict that the minimized lattice thermal conductivity $\kappa\sub{L}$ occurs near the equiatomic \ch{Pb_{0.5}Sn_{0.5}Te_{0.5}Se_{0.5}} composition, correlating as expected with the region of maximal compositional disorder (see Figure \ref{fig:quat_chalc_kL}). In contrast to the previous half-Heusler examples, this reduction is achieved by alloying on different sublattices, in this case the cation and anion sites. The form of the Klemens model suggests that point defect substitutions on different sublattices should scatter independently such that an improvement from combining these orthogonal effects could be expected. A possible justification for this orthogonality is that different sublattices contribute very differently to the overall lattice vibrations of the structure, for example more massive sublattices will dominate the low-frequency range. The Tamura model\cite{Tamura1984} for phonon-impurity scattering further elucidates the separation in targeted phonon frequency window for sublattices of different mass. In Supplementary Section \ref{supp:tamura}, we illustrate the separation in scattering frequency window for the \ch{Na} and \ch{Cl} sublattices comprising \ch{NaCl}.

\begin{figure}[h!]
    \centering
    \includegraphics[width=\textwidth]{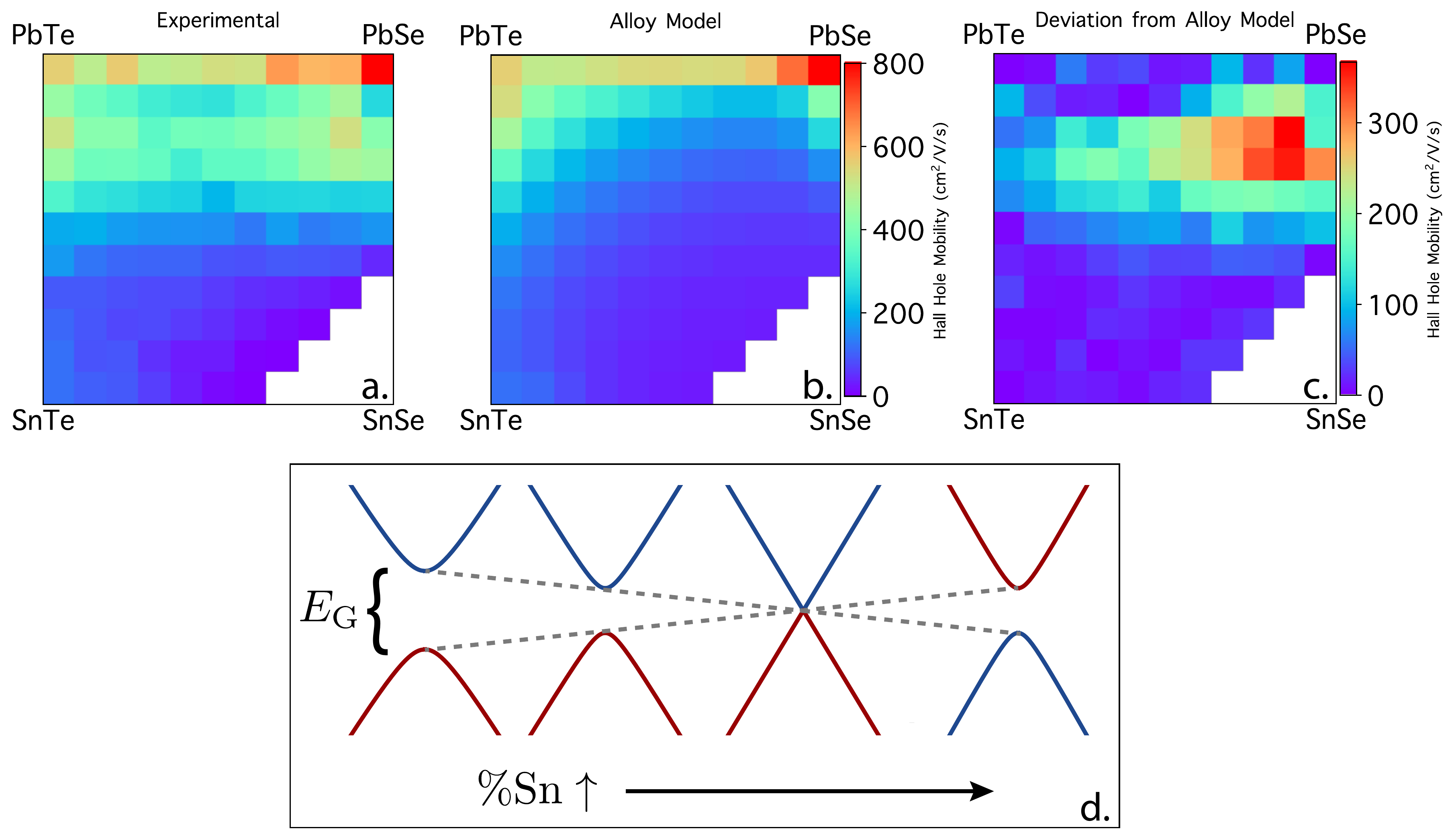}
    \caption{Hall mobility heatmap in p-type quaternary PbTe-PbSe-SnTe-SnSe alloy system. The experimental data (a) is reproduced from Ortiz \textit{et al.}\cite{Ortiz2019} and is compared to the alloy mobility model extrapolated using the Muggianu method (b). The deviation between the experiment and mobility $\mu\sub{exp} - \mu\sub{alloy}$ in also plotted (c), which helps to isolate effective mass changes in the alloy regime. Panel (c) shows a horizontal band of high observed mobility at about 20-30\% Sn content, which may be attributable to a low hole effective mass. This region corresponds to a documented band inversion along the PbTe-SnTe\cite{Arachchige2009, Ortiz2019, Dimmock1966, Dmitriev1986} and PbSe-SnSe\cite{Wang2019} binary systems. Prior to the band inversion, the band gap narrows while the conduction and valence bands linearize according to a Kane band dispersion model\cite{Pei2012}, producing the region of low effective mass.}
    \label{fig:quat_hole_mobility}
\end{figure}

We additionally applied an alloy scattering model to the Hall mobility data for this p-type material. Visual inspection of the experimental data, alone, suggests that bandstructure changes independent of alloy scattering are likely at play. Rather than following the expected U-shaped curve behavior predicted by Nordheim's rule, there is a horizontal band of relatively high mobility at about 20-30\% Sn content that corresponds in location to a documented band inversion along both the PbTe-SnTe\cite{Arachchige2009, Dimmock1966, Dmitriev1986} and PbSe-SnSe\cite{Wang2019} binary alloy systems. In the vicinity of the band inversion, the band gap shrinks and the valence and conduction bands linearize according to a Kane dispersion model, leading to a near zero effective mass that produces the peak in hole mobility (see Figure \ref{fig:quat_hole_mobility}d). Such bandstructure changes are not captured by the perturbation theory alloy scattering model, and by plotting the deviation between experimental and model mobility, we see the effective mass change highlighted around 30\% Sn content in Figure \ref{fig:quat_hole_mobility}c. 

There is also a reduction in mobility at about 50\% Sn content, which is likely due to an increase in doping level near the SnTe-SnSe side of the quaternary alloy that leads to a sampling of the second, heavier-mass $\Sigma$-pocket valence band. The alloy scattering potential is, therefore, expected to change in the vicinity of different end-member compounds. In the alloy mobility model fit to the data along the pseudobinaries, we allow the alloy scattering potential $\Delta U$ to vary with alloy composition as a Redlich-Kister polynomial up to the degree $n = 2$ in order to adequately describe the experimental data. The Muggianu method is then applied to extrapolate from the pseudobinary to pseudoquaternary space.

\subsection{Multicomponent Alloy Design Rules}

The point defect perturbation theory approach to alloy thermal conductivity points to scenarios in which multicomponent alloying can be beneficial over a simple binary alloy. As demonstrated in several of the pseudoternary half-Heusler alloy systems, increasing configurational entropy by forming equiatomic multicomponent alloys is not always an effective strategy to reduce the thermal conductivity. In fact, introducing an alloy element of intermediate atomic mass and radius can reduce the overall mass and strain contrast scattering. 

\begin{figure}
    \centering
    \includegraphics[width = 0.8\textwidth]{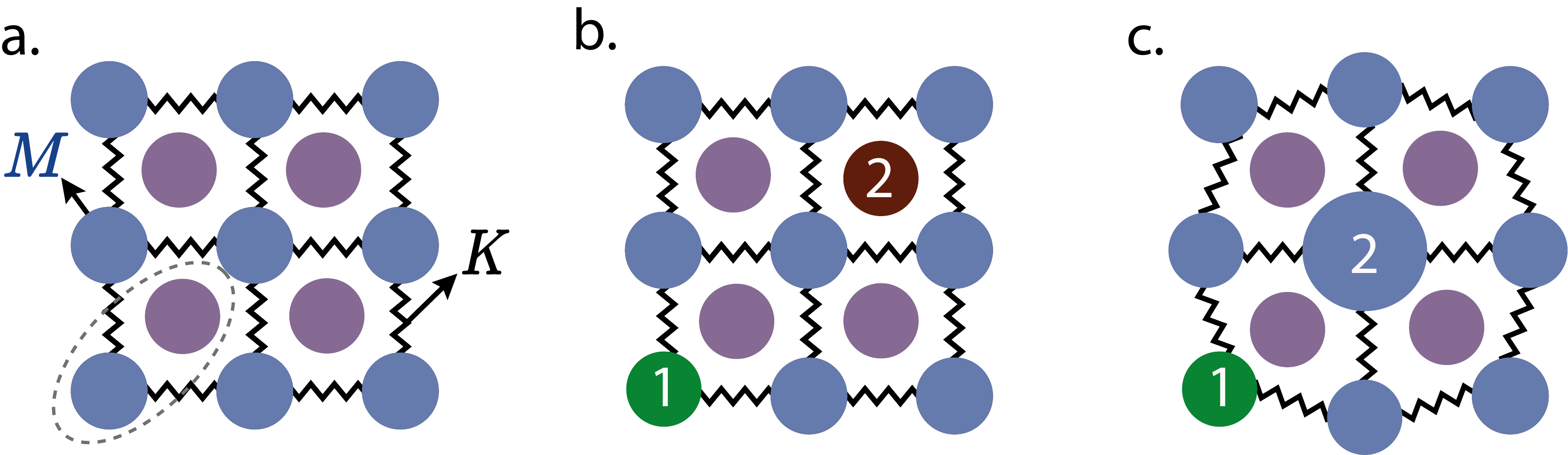}
    \caption{Multicomponent alloy design strategies for reduced thermal conductivity due to point defect scattering should take advantage of orthogonal scattering effects. (a) Schematic of lattice with 2-atom primitive unit cell basis (encircled by dotted line). (b) Alloy elements, labelled 1 and 2, substitute on different sublattices. (c) Alloy element 1 contributes significant mass contrast while alloy element 2 contributes significant strain contrast.}
    \label{fig:multicomponent_alloy_strategies}
\end{figure}

Instead, the additional alloy element should introduce an orthogonal scattering effect, i.e. a scattering effect that, to first order, acts independently of the ones already at play in the lower dimensional alloy system. Within the realm of point defect scattering, the following strategies exist (depicted in Figure \ref{fig:multicomponent_alloy_strategies}):

\begin{enumerate}
    \item Alloy on different sublattices of the compound(e.g. \ch{Pb_{1-x}Sn_xTe_{1-y}Se_y})
    \item Use separate alloy elements to introduce mass and strain contrast into the system
\end{enumerate}

The first strategy is demonstrated in the quaternary chalcogenide system discussed earlier, where the cation and anion site are both alloyed, and the minimum thermal conductivity occurs at the equiatomic \ch{Pb_{0.5}Sn_{0.5}Te_{0.5}Se_{0.5}} composition. The same multi-site alloying strategy may explain the origin of the lattice thermal conductivity reduction in the recently reported high-performance PbSe-based multicomponent thermoelectric alloy\cite{Jiang2021}.

The latter strategy is likely more exotic, since atomic mass and size are often correlated within the same coordination environment. However, if mass and strain contrast are each maximized along different $n$-nary alloy systems, it is likely that entering  an $n+1$-nary system will lead to a region of further suppressed thermal conductivity. The orthogonality between mass and strain scattering stems from the fact that the two phenomena effect different terms of the lattice energy, mass variance perturbs the kinetic term while the strain contrast perturbs the potential energy. Supplementary Section \ref{suppsct:toy_alloy} shows a simple demonstration of this phenomenon in a toy alloy model composed of fictitious atoms $A$, $B$, and $C$, in which the atomic mass trend is $A< B = (A + C)/2 <C$ and the atomic radius trend is $B< C = (A+B)/2 <A$. With mass contrast maximized along the $A-B$ binary and radius contrast maximized along the $A-C$ binary, the minimum $\kappa\sub{L}$ value is observed near the center of the ternary diagram.  

While the first two mechanisms are consistent with phonon--point-defect scattering, phonons are also affected by dislocations and phase boundaries, which can be tailored through alloying. Dislocation-phonon interactions have been shown to explain the high thermoelectric performance in numerous materials, as the lattice strain produced from high dislocation densities will lead to strong lattice softening and phonon scattering effects\cite{Hanus2019, Male2021_mechanical, Wu2020, Sun2021}. Alloying elements can immobilize dislocations, preventing dislocation glide and annihilation, in order to maintain higher dislocation densities\cite{Male2021, Wu2019, Kim2016, Yu2018}. Since the perturbation to elastic constants in addition to the elastic strain around the defect is primarily responsible for this dislocation pinning, point defects with a high $\Gamma\sub{R}$ parameter would be most effective for this strategy. In the case of \ch{PbTe}, co-doping with \ch{Na} and \ch{Eu} appears to best maintain a high dislocation density even when dislocations become more mobile at elevated temperatures, performing better than samples with just a single dopant\cite{Male2021, Wu2019}. This suggests, then, that multiple defects may better preserve dislocation strain. 

Finally, alloying to achieve microstructural changes, through the formation of grain boundaries or even secondary phases can be an effective route to reduced thermal conductivity\cite{Appel2015, Schwall2013}. For example, introducing \ch{MnTe} into both the \ch{PbTe-SrTe} alloy as well as the \ch{PbTe-SnTe-GeTe} alloy was shown to produce low-angle grain boundaries and precipitate boundaries whose interfacial dislocation arrays effectively scattered mid-frequency phonons to suppress the thermal conductivity\cite{Luo2017, Hu2018}.

\section{Conclusion}

Multicomponent alloys formed between three or more thermoelectric compounds are an uncharted territory of high interest, primarily because of the large degrees of compositional disorder which should be effective in scattering heat-carrying phonons. Here, we show an extension of virtual crystal approximation alloy scattering models to higher order alloy systems. We show for both electronic and thermal transport that the scattering parameter $\Gamma$ can be written as a sum over ``binary terms," those which depend only on two components in the system. Their functional form mimics that of excess Gibbs free energy, allowing for an analogy to be made to computational thermodynamics literature. We are therefore able to apply the Redlich-Kister polynomial and Muggianu method to better extrapolate alloy scattering models to higher dimensions. Our modelling indicates that additional alloying and introduction of configurational entropy does not, by necessity, reduce the thermal conductivity. To produce a thermal conductivity reduction through multicomponent alloying, the additional alloying element should introduce an orthogonal form of scattering. Within the realm of point defect scattering, this may include combining alloy elements that occupy different sublattices or using separate alloy elements to introduce mass and strain contrast into the system. Additionally, seeking microstructure changes via alloying, through modifications of dislocation density for example, is also a viable strategy to achieve a lattice thermal conductivity reduction. Given that multicomponent alloys are currently being investigated for a wide class of energy materials, we hope that these analytic alloy models can be applied to a wide range of transport-related phenomena beyond thermoelectric research.

\section{Data and Code Availability}

The literature data used to perform the analyses of the half-Heusler and IV-VI semiconductor alloy systems are tabulated in the Supplementary Material. Python scripts used to implement model can be found at \texttt{https://github.com/RamyaGuru/MulticomponentTE}.  

\section{Acknowledgements}

We acknowledge support from the U.S. Department of Energy, Office of Energy Efficiency and Renewable Energy (EERE) program ``Accelerated Discovery of Compositionally Complex Alloys for Direct Thermal Energy Conversion" (DOE Award DE-AC02-76SF00515). RG and GJS additionally acknowledge the following financial
assistance award 70NANB19H005 from U.S.
Department of Commerce, National Institute of
Standards and Technology as part of the Center for
Hierarchical Materials Design (CHiMaD).

\beginsupplement

% \section{Interpreting Deviations from Alloy Mobility Model}
% The electron scattering rate can be computed using 

\section{Pseudoternary half-Heusler Experimental Data}\label{suppsct:hh_exp_data}

The experimental transport data tabulated here was collected, in part, using the \texttt{StarryData2} database\cite{Katsura2019}.

% Table generated by Excel2LaTeX from sheet 'LatexSheet'
% Table generated by Excel2LaTeX from sheet 'LatexSheet'
\begin{center}
\begin{longtable}{|l|l|l|l|l|l|l|}
  \caption{\textbf{Experimental Transport Function ($\sigma\sub{E0}$) and Lattice Thermal Conductivity ($\kappa\sub{L}$) Values from Literature for the \textit{X}NiSn System} The $\sigma\sub{E0}$ values were calculated using the approach described in Section \ref{subsct:background_sigmae0} when $S$ and $\sigma$ pairs were available for samples at multiple doping levels. The $\kappa\sub{L}$ values were calculated by subtracting off the electronic thermal conductivity term modeled using the Wiedemann-Franz law.}\\
%    \begin{tabular}{rrrrrrlr}
    \hline 
    \multicolumn{1}{|c|}{x(TiNiSn)} &
    \multicolumn{1}{c|}{x(ZrNiSn)} &
    \multicolumn{1}{c|}{x(HfNiSn)} &
    \multicolumn{1}{c|}{$\sigma_{\mathrm{E0}}$ (S/m)} &
    \multicolumn{1}{c|}{$\kappa_{\mathrm{L}}$ (W/m/K)} & 
    \multicolumn{1}{c|}{Dopant} & 
    \multicolumn{1}{c|}{Source}  \\
    \hline
   \endfirsthead
    
    \hline
    \multicolumn{7}{|c|}%
    {{\bfseries \tablename\ \thetable{} -- continued from previous page}} \\
    \hline
    \multicolumn{1}{|c|}{x(TiNiSn)} &
    \multicolumn{1}{c|}{x(ZrNiSn)} &
    \multicolumn{1}{c|}{x(HfNiSn)} &
    \multicolumn{1}{c|}{$\sigma_{\mathrm{E0}}$ (S/m)} &
    \multicolumn{1}{c|}{$\kappa_{\mathrm{L}}$ (W/m/K)} & 
    \multicolumn{1}{c|}{Dopant} & 
    \multicolumn{1}{c|}{Source}  \\
    \hline
    \endhead
    
    \hline \multicolumn{7}{|r|}{{Continued on next page}} \\ \hline
    \endfoot

    \hline \hline
    \endlastfoot

    \hline
    % x(TiNiSn) & x(ZrNiSn) & x(HfNiSn) & $\sigma_{\mathrm{E0}}$ (S/m) & $\kappa_{\mathrm{L}}$ (W/m/K) & Dopant & \multicolumn{1}{r}{Source} &  \\
    0     & 0.25  & 0.75  & 70125 & 5.38  & Sb    & \cite{Akram2016} \\
    0.3   & 0.35  & 0.35  &       & 3.88  & Bi    & \cite{Appel2015} \\
    0.3   & 0.35  & 0.35  &       & 3.66  & excess Ni & \cite{Appel2014} \\
    0     & 0     & 1     &       & 6.38  &       & \cite{Schrade2017} \\
    0     & 1     & 0     &       & 6.74  &       & \cite{Schrade2017} \\
    1     & 0     & 0     &       & 3.85  &       & \cite{Schrade2017} \\
    0.5   & 0.5   & 0     &       & 3.31  &       & \cite{Schrade2017} \\
    0.5   & 0     & 0.5   &       & 2.45  &       & \cite{Schrade2017} \\
    0     & 0.5   & 0.5   &       & 2.3   &       & \cite{Schrade2017} \\
    0.5   & 0.25  & 0.25  &       & 2.83  &       & \cite{Schrade2017} \\
    0.5   & 0.25  & 0.25  & 60937 & 2.91 & Sb    & \cite{Fan2014} \\
    0.25  & 0.75  & 0     &       & 3.07 &       & \cite{Gurth2016} \\
    0.5   & 0.5   & 0     &       & 2.42 &       & \cite{Gurth2016}\\
    0.75  & 0.25  & 0     &       & 3.73 &       & \cite{Gurth2016} \\
    1     & 0     & 0     &       & 3.18 &       & \cite{Gurth2016} \\
    0     & 1     & 0     &       & 3.49 &       & \cite{Gurth2016} \\
    0.25  & 0     & 0.75  &       & 2.87 &       & \cite{Gurth2016}\\
    0.5   & 0     & 0.5   &       & 2.75 &       & \cite{Gurth2016} \\
    0.75  & 0     & 0.25  &       & 2.48 &       & \cite{Gurth2016} \\
    0     & 0     & 1     &       & 3.46 &       & \cite{Gurth2016} \\
    0     & 0.75  & 0.25  &       & 4.13  &       & \cite{Gurth2016} \\
    1     & 0     & 0     & 84677 & 6.95  & Sb    & \cite{Kim2007} \\
    0.95  & 0     & 0.05  & 116606 & 5.21  & Sb    & \cite{Kim2007} \\
    0.8   & 0     & 0.2   & 76209 & 4.94  & Sb    & \cite{Kim2007} \\
    0.3   & 0.35  & 0.35  &       & 3.58  &       & \cite{Kurosaki2005} \\
    0     & 1     & 0     &       & 12.72 &       & \cite{Muta2005} \\
    0.3   & 0.7   & 0     &       & 7.39  &       & \cite{Muta2005} \\
    0.5   & 0.5   & 0     &       & 5.30 &       & \cite{Muta2005} \\
    0.5   & 0.5   & 0     &       & 4.67 &       & \cite{Muta2005} \\
    0.33  & 0.33  & 0.33  &       & 5.67  &       & \cite{Populoh2013} \\
    0     & 0.5   & 0.5   &       & 2.81  &       & \cite{Populoh2013} \\
    0     & 0.5   & 0.5   &       & 4.16  &       & \cite{Populoh2013} \\
    0     & 0.5   & 0.5   &       & 3.99  &       & \cite{Sakurada2005} \\
    0.2   & 0.4   & 0.4   &       & 3.74  &       & \cite{Sakurada2005} \\
    0.3   & 0.35  & 0.35  &       & 3.09  &       & \cite{Sakurada2005} \\
    0.5   & 0.25  & 0.25  & 158357 & 2.92  & Sb    & \cite{Sakurada2005} \\
    0.7   & 0.15  & 0.15  &       & 3.3   &       & \cite{Sakurada2005} \\
    0.5   & 0.25  & 0.25  &       & 3.3   &       & \cite{Tang2009} \\
    0     & 0.75  & 0.25  &       & 3.61  &       & \cite{Yu2012} \\
    0     & 0.4   & 0.6   & 85158 & 3.3   & Sb    & \cite{Yu2009} \\
    0     & 0.3   & 0.7   & 72348 & 4     & Sb    & \cite{Yu2009} \\
    0.5   & 0.5   & 0     & 63378 & 4     & Sb    & \cite{Downie2014} \\
    0     & 0.9   & 0.1   & 107439 & 3.3   & Sb    & \cite{Chen2013} \\
    0     & 0.2   & 0.8   & 93747 & 4.5   & Sb    & \cite{Liu2015} \\
    0     & 0     & 1     & 85915 & 6.5   & Sb    & \cite{Liu2015} \\
    0.5   & 0.25  & 0.25  & 43947 & 3.58  & Sb    & \cite{Schwall2013}  \\
    0.43  & 0.28  & 0.29  & 35671 & 5.6   &       & \cite{SchwallThesis} \\
    0.21  & 0.4   & 0.39  & 50008 & 9.1   &       & \cite{SchwallThesis}
%    \end{tabular}%
  \label{tab:XNiSn_data}%
\end{longtable}%
\end{center}

% Table generated by Excel2LaTeX from sheet 'Sheet7'
%Remove multiphase TECCA samples

\begin{center}
\begin{longtable}{|l|l|l|l|l|l|l|}
  \caption{\textbf{Experimental Transport Function ($\sigma\sub{E0}$) and Lattice Thermal Conductivity ($\kappa\sub{L}$) Values from Literature for the \textit{X}FeSb System} All values are reported for samples with 20\% Ti doping. The $\kappa\sub{L}$ values were calculated by subtracting off the electronic thermal conductivity term modeled using the Wiedemann-Franz law.}\\
    \hline 
    \multicolumn{1}{|c|}{x(VFeSb)} &
    \multicolumn{1}{c|}{x(NbFeSb)} &
    \multicolumn{1}{c|}{x(TaFeSb)} &
    \multicolumn{1}{c|}{$\sigma_{\mathrm{E0}}$ (S/m)} &
    \multicolumn{1}{c|}{$\kappa_{\mathrm{L}}$ (W/m/K)} & 
    \multicolumn{1}{c|}{Dopant} & 
    \multicolumn{1}{c|}{Source}  \\
    \hline
   \endfirsthead
    
    \hline
    \multicolumn{7}{|c|}%
    {{\bfseries \tablename\ \thetable{} -- continued from previous page}} \\
    \hline
    \multicolumn{1}{|c|}{x(VFeSb)} &
    \multicolumn{1}{c|}{x(NbFeSb)} &
    \multicolumn{1}{c|}{x(TaFeSb)} &
    \multicolumn{1}{c|}{$\sigma_{\mathrm{E0}}$ (S/m)} &
    \multicolumn{1}{c|}{$\kappa_{\mathrm{L}}$ (W/m/K)} & 
    \multicolumn{1}{c|}{Dopant} & 
    \multicolumn{1}{c|}{Source}  \\
    \hline
    \endhead
    
    \hline \multicolumn{7}{|r|}{{Continued on next page}} \\ \hline
    \endfoot

    \hline \hline
    \endlastfoot

    \hline
    0.05  & 0.95  & 0     & 120353 & 4.85  & Ti    & \cite{Fu2016AEM} \\
    0.1   & 0.9   & 0     & 129429 & 4.52  & Ti    & \cite{Fu2016AEM} \\
    0.25  & 0.75  & 0     & 115760 & 3.69  & Ti    & \cite{Fu2016AEM} \\
    0.4   & 0.6   & 0     & 97073 & 3.07 & Ti    & \cite{Fu2016AEM} \\
    0.55  & 0.45  & 0     & 92508 & 3.28  & Ti    & \cite{Fu2016AEM} \\
    0.7   & 0.3   & 0     & 81242 & 3.18  & Ti    & \cite{Fu2016AEM} \\
    0.85  & 0.15  & 0     & 76770 & 3.44  & Ti    & \cite{Fu2016AEM} \\
    1     & 0     & 0     & 47298 & 3.17  & Ti    & \cite{Fu2016AEM} \\
    0     & 1     & 0     & 130483 & 4.45  & Ti    & \cite{Fu2016AEM} \\
    0     & 0.96  & 0.04  & 666753 & 3.99 & Ti    & \cite{Yu2017AEM} \\
    0     & 0.88  & 0.12  & 540480 & 3.47 & Ti    & \cite{Yu2017AEM} \\
    0     & 0.8   & 0.2   & 550582 & 2.95 & Ti    & \cite{Yu2017AEM} \\
    0     & 0.76  & 0.24  & 651593 & 2.12 & Ti    & \cite{Yu2017AEM} \\
    0     & 0.68  & 0.32  & 572491 & 2.39 & Ti    & \cite{Yu2017AEM} \\
    0     & 0.64  & 0.36  & 548937 & 2.13 & Ti    & \cite{Yu2017AEM} \\
    0     & 0.6   & 0.4   & 538820 & 1.76 & Ti    & \cite{Yu2017AEM} \\
    0.05  & 0     & 0.95  & 138242 & 2.52  & Ti    & \cite{Zhu2019NatComm} \\
    0.1   & 0     & 0.9   & 177881 & 2.33  & Ti    & \cite{Zhu2019NatComm} \\
    0.15  & 0     & 0.85  & 247795 & 2.46  & Ti    & \cite{Zhu2019NatComm} \\
    0.1875 & 0     & 0.8125 & 74954 & 1.22 & Ti    & unpublished \\
    0.1   & 0.7   & 0.2   & 87960 & 0.95 & Ti    & unpublished \\
    0.63  & 0     & 0.37  & 53170 & 3.26 & Ti    & unpublished \\
    0.34125 & 0.325 & 0.33375 & 44879& 2.03 & Ti    & unpublished \\
    0     & 0.8   & 0.2   & 121976 & 3.65 & Ti    & unpublished \\
    0     & 0.5   & 0.5   & 136446 & 3.64 & Ti    & unpublished \\
    0     & 0.2   & 0.8   & 82195 & 3.33 & Ti    & unpublished \\
    0.6   & 0.2   & 0.2   & 41556 & 2.54 & Ti    & unpublished \\
    0 & 0 & 1     & 151851 & 5.5   & Ti    & unpublished \\
    0 & 1     & 0 & 143928 & 7.42  & Ti    & unpublished
  \label{tab:XFeSb_data}%
\end{longtable}%
\end{center}

% Table generated by Excel2LaTeX from sheet 'Sheet8'
\begin{center}
\begin{longtable}{|l|l|l|l|l|l|l|}
  \caption{\textbf{Experimental Transport Function ($\sigma\sub{E0}$) and Lattice Thermal Conductivity ($\kappa\sub{L}$) Values from Literature for the \textit{X}CoSb System} The $\sigma\sub{E0}$ values were calculated using the approach described in Section \ref{subsct:background_sigmae0} when $S$ and $\sigma$ pairs were available for samples at multiple doping levels. The $\kappa\sub{L}$ values were calculated by subtracting off the electronic thermal conductivity term modeled using the Wiedemann-Franz law.}\\
    \hline 
    \multicolumn{1}{|c|}{x(TiCoSb)} &
    \multicolumn{1}{c|}{x(ZrCoSb)} &
    \multicolumn{1}{c|}{x(HfCoSb)} &
    \multicolumn{1}{c|}{$\sigma_{\mathrm{E0}}$ (S/m)} &
    \multicolumn{1}{c|}{$\kappa_{\mathrm{L}}$ (W/m/K)} & 
    \multicolumn{1}{c|}{Dopant} & 
    \multicolumn{1}{c|}{Source}  \\
    \hline
   \endfirsthead
    
    \hline
    \multicolumn{7}{|c|}%
    {{\bfseries \tablename\ \thetable{} -- continued from previous page}} \\
    \hline
    \multicolumn{1}{|c|}{x(TiCoSb)} &
    \multicolumn{1}{c|}{x(ZrCoSb)} &
    \multicolumn{1}{c|}{x(HfCoSb)} &
    \multicolumn{1}{c|}{$\sigma_{\mathrm{E0}}$ (S/m)} &
    \multicolumn{1}{c|}{$\kappa_{\mathrm{L}}$ (W/m/K)} & 
    \multicolumn{1}{c|}{Dopant} & 
    \multicolumn{1}{c|}{Source}  \\
    \hline
    \endhead
    
    \hline \multicolumn{7}{|r|}{{Continued on next page}} \\ \hline
    \endfoot

    \hline \hline
    \endlastfoot
    
    0     & 0     & 1     & 84928 & 4.39 & Sb    & \cite{Rausch2014} \\
    0     & 0.5   & 0.5   & 73550 & 2.64 & Sb    & \cite{Rausch2014} \\
    0.5   & 0     & 0.5   & 68372 & 2.33 & Sb    & \cite{Rausch2014} \\
    0.5   & 0.5   & 0     & 55161 & 4.08 & Sb    & \cite{Rausch2014} \\
    0     & 1     & 0     & 70354 & 4.39 & Sb    & \cite{Rausch2014} \\
    1     & 0     & 0     & 56680 & 6.46 & Sb    & \cite{Rausch2014} \\
    0.3   & 0.35  & 0.35  & 68706 & 2.4   & Sb    & \cite{Rausch2015} \\
    1     & 0     & 0     &       & 14.7  &       & \cite{Qiu2009} \\
    0.9   & 0.1   & 0     &       & 10.9  &       & \cite{Qiu2009} \\
    0.8   & 0.2   & 0     &       & 7.64  &       & \cite{Qiu2009} \\
    0.6   & 0.4   & 0     & 46205 & 6.97  & Ni    & \cite{Qiu2009} \\
    0.5   & 0.5   & 0     &       & 7.6   &       & \cite{Qiu2009} \\
    0     & 1     & 0     &       & 15.6  &       & \cite{Silpawilawan2017} \\
    0     & 0     & 1     &       & 11.9  &       & \cite{Sekimoto2005}
  \label{tab:XCoSb_data}%
\end{longtable}%
\end{center}

%Table generated by Excel2LaTeX from sheet 'Sheet9'
\begin{center}
\begin{longtable}{|l|l|l|l|l|l|l|}
  \caption{\textbf{Experimental Transport Function ($\sigma\sub{E0}$) and Lattice Thermal Conductivity ($\kappa\sub{L}$) Values from Literature for the \textit{X}CoSn System} The $\sigma\sub{E0}$ values were calculated using the approach described in Section \ref{subsct:background_sigmae0} when $S$ and $\sigma$ pairs were available for samples at multiple doping levels. The $\kappa\sub{L}$ values were calculated by subtracting off the electronic thermal conductivity term modeled using the Wiedemann-Franz law.}\\
    \hline 
    \multicolumn{1}{|c|}{x(VCoSn)} &
    \multicolumn{1}{c|}{x(NbCoSn)} &
    \multicolumn{1}{c|}{x(TaCoSn)} &
    \multicolumn{1}{c|}{$\sigma_{\mathrm{E0}}$ (S/m)} &
    \multicolumn{1}{c|}{$\kappa_{\mathrm{L}}$ (W/m/K)} & 
    \multicolumn{1}{c|}{Dopant} & 
    \multicolumn{1}{c|}{Source}  \\
    \hline
   \endfirsthead
    
    \hline
    \multicolumn{7}{|c|}%
    {{\bfseries \tablename\ \thetable{} -- continued from previous page}} \\
    \hline
    \multicolumn{1}{|c|}{x(VCoSn)} &
    \multicolumn{1}{c|}{x(NbCoSn)} &
    \multicolumn{1}{c|}{x(TaCoSn)} &
    \multicolumn{1}{c|}{$\sigma_{\mathrm{E0}}$ (S/m)} &
    \multicolumn{1}{c|}{$\kappa_{\mathrm{L}}$ (W/m/K)} & 
    \multicolumn{1}{c|}{Dopant} & 
    \multicolumn{1}{c|}{Source}  \\
    \hline
    \endhead
    
    \hline \multicolumn{7}{|r|}{{Continued on next page}} \\ \hline
    \endfoot

    \hline \hline
    \endlastfoot
    0     & 0     & 1     & 38603 & 5.94  & Sb    & \cite{Li2019ACSAMI} \\
    0     & 1     & 0     & 91932 & 7.72  & Sb    & \cite{Li2019ACSAMI} \\
    0     & 0.4   & 0.6   & 56619 & 4.8   & Sb    & \cite{Li2019ACSAMI} \\
    1     & 0     & 0     & 16316 & 12.8  &       & \cite{Zaferani2020}
  \label{tab:XCoSn_data}%
\end{longtable}%
\end{center}

\section{Pseduoternary half-Heusler Alloy Model Predictions}

\begin{figure}
    \centering
    \includegraphics[width = 0.6\textwidth]{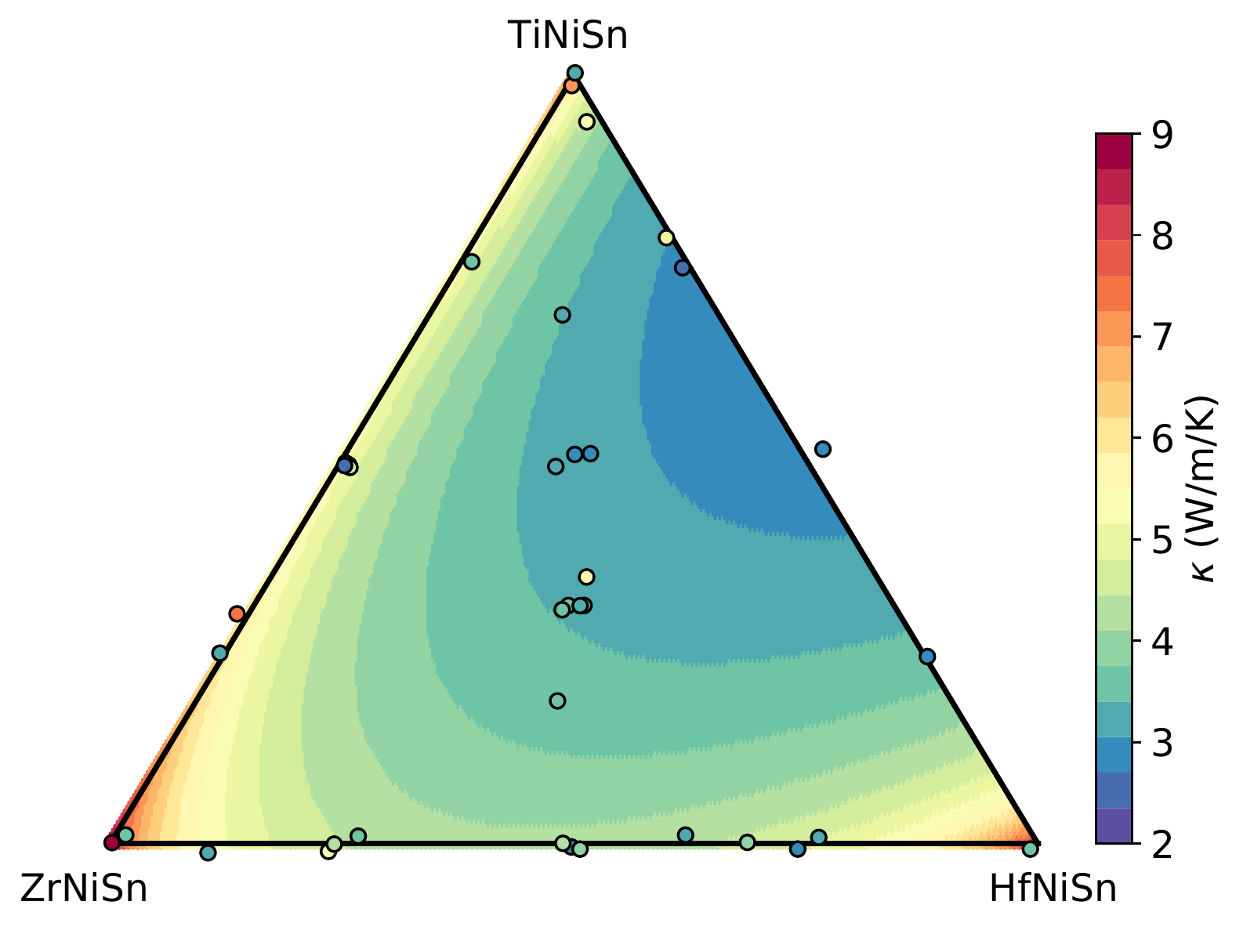}
    \caption{\textit{X}NiSn ternary plot of the alloy model lattice thermal conductivity $\kappa\sub{L}$ with the full collected literature data summarized in Table \ref{tab:XNiSn_data}. A small amount of random noise is applied to the composition to  better view overlapping datapoints.  A large spread in values is observed for samples of  the same composition, which may be attributable to variations in microstructure such as grain size.}
    \label{fig:XNiSn_kL_full_data}
\end{figure}

\begin{figure}
    \centering
    \includegraphics[width=\textwidth]{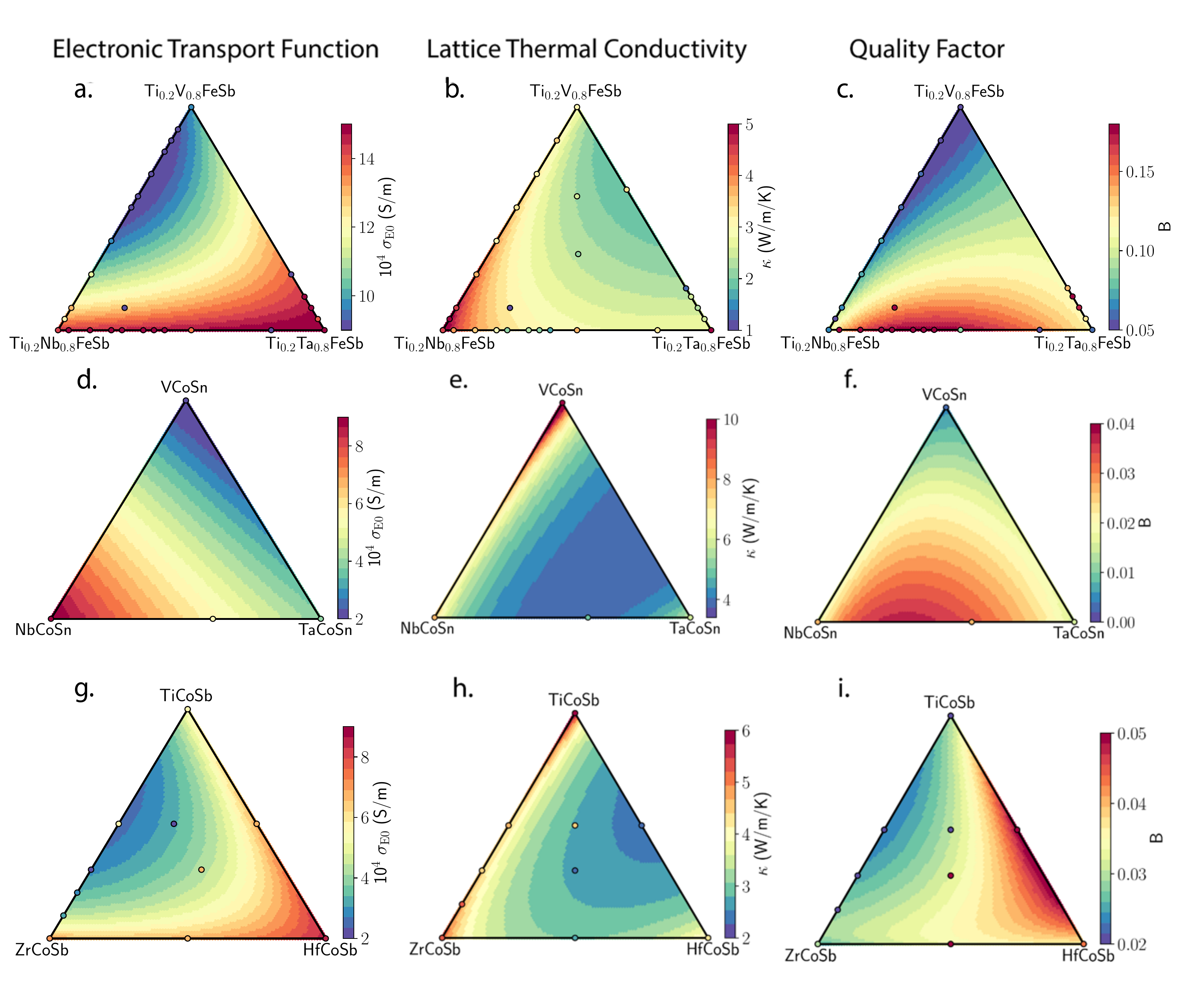}
    \caption{Alloy model predictions with overlaid experimental scatter points for three compound families: $X$FeSb (with 20\% Ti doping), $X$CoSn, and $X$CoSb. For each system, the electronic transport function $\sigma\sub{E0}$, lattice thermal conductivity $\kappa\sub{L}$, and quality factor $B$ are shown. Although each compound family has the same motif for lattice thermal conductivity, with $\kappa\sub{L}$ minimized along the binary with highest mass contrast, the $B$ factor plots show very different patterns of high and low performance regions. This speaks to the trade-off  between thermal and electronic property variation with alloying.}
    \label{fig:hh_tern_grid}
\end{figure}

\section{Alloy System Model Parameters}

\begin{table}[htbp]
  \centering
  \caption{End-Member lattice properties for half-Heusler pseudoternary systems used in alloy scattering models for phonons and charge carriers}
    \begin{tabular}{|p{3 cm}|p{3 cm}|p{3 cm}|p{3 cm}|p{2 cm}|}
    \hline
    \textbf{Composition} & \textbf{Speed of Sound}\newline (m/s) & \textbf{Volume Per}\newline \textbf{Atom} ($\AA^3$) & \textbf{Effective Mass (e or h)} & \textbf{Source} \\
    \hline
    VFeSb & 2374  & 16.2  & 2.5(h) & \cite{Fu2012}, \cite{Page2016} \\
    \hline
    NbFeSb & 3052  & 17.7  & 1.6(h) & \cite{Silpawilawan2017}, \cite{Page2016} \\
    \hline
    TaFeSb & 3056  & 17.4  & 1.65(h) & \cite{Grytsiv2019} \\
    \hline
    VCoSn & 4760  & 17.4  &       & \cite{DeJong2015} \\
    \hline
    NbCoSn & 3368  & 17.5  &       & \cite{Li2019ACSAMI}\\
    \hline
    TaCoSn & 3077  & 17.5  &       & \cite{Li2019ACSAMI}\\
    \hline
    TiNiSn & 3553  & 17.5  & 0.56(e), 0.39(h) & \cite{Gandi2016} \\
    \hline
    ZrNiSn & 2914  & 19.3  & 0.38(e), 0.25(h) & \cite{Gandi2016} \\
    \hline
    HfNiSn & 2756  & 19.1  & 0.36(e), 0.22(h) & \cite{Gandi2016} \\
    \hline
    TiCoSb & 3592 & 17    &       & \cite{Sekimoto2005} \\
    \hline
    ZrCoSb & 3490  & 18.6  &       & \cite{Sekimoto2005}, \cite{Silpawilawan2017} \\
    \hline
    HfCoSb & 3012  & 18.3  &       & \cite{Sekimoto2005} \\
    \hline
    \end{tabular}%
  \label{tab:hhtern_model_param}%
\end{table}%

% Table generated by Excel2LaTeX from sheet 'Sheet2'
\begin{table}[htbp]
  \centering
  \caption{Coefficients for the Redlich-Kister polynomial representations of the scattering parameter $\Gamma$ in both the electronic and thermal transport models. Coefficients are designated as ${}^kD$, where $k$ is the order of the term in the polynomial.}
    \begin{tabular}{|p{3 cm}|p{3 cm}|p{3 cm}|p{3 cm}|}
    \multicolumn{4}{l}{\textbf{Carrier Mobility Model: $\mathbf{\Gamma\sub{el}}$ Expression}}\\
    \hline
    \textbf{Binary System} & $\mathbf{{}^0D}$ & $\mathbf{{}^1D}$ & $\mathbf{{}^2D}$ \\
    \hline
    PbSe-SnSe & 0     & -15.8 & 15.8 \\
    \hline
    SnTe-SnSe & 5.3   & -17.6 & 12 \\
    \hline
    PbTe-SnSe & 0     & -12.2 & 8.4 \\
    \hline
    SnTe-PbSe & 1.22  & 11    & 14.9 \\
    \hline
    PbTe-PbSe & 1.48  & -2    & 3.2 \\
    \hline
    PbTe-SnTe & 0     & -7.5  & 6.5 \\
    \hline
    \multicolumn{4}{l}{\textbf{Lattice Thermal Conductivity Model: $\mathbf{\Gamma\sub{R}}$ Expression}}\\
    \hline
    \textbf{Binary System} & $\mathbf{{}^0D}$ & $\mathbf{{}^1D}$ & $\mathbf{{}^2D}$ \\
    \hline
    PbSe-SnSe & 8.1   & 0     & 8.5 \\
    \hline
    SnTe-SnSe & 0.73  & 0     & 2.2 \\
    \hline
    PbTe-SnSe & 3.3   & 3.8   & 1.1 \\
    \hline
    SnTe-PbSe & 4.4   & 9     & 0 \\
    \hline
    PbTe-PbSe & 2.6   & 1.6   & 0 \\
    \hline
    PbTe-SnTe & 94    & 96    & 0 \\
    \hline
    \end{tabular}%
  \label{tab:hhtern_rk_param}%
\end{table}%

% \begin{itemize}
%     \item Table for the quaternary thermal conductivity and mobility Redlich-Kister polynomials
%     \item Table for the end-member half-Heusler data (non-transport)
% \end{itemize}

\section{Decoupled Mass and Strain Scattering}\label{suppsct:toy_alloy}

In this section, we will consider a fictitious alloy with components $A$, $B$, and $C$. Each component has an equivalent pristine lattice thermal conductivity, speed of sound, and lattice parameter. The relationship between their atomic mass is: $A< B = (A + C)/2 <C$, such that the highest mass contrast exists between the $A-C$ binary. While the relationship between atomic radius is: $B< C = (A+B)/2 <A$, such that the highest strain contrast exists between the $A-B$ binary. 

\begin{figure}[h!]
    \centering
    \includegraphics[width = 0.8\textwidth]{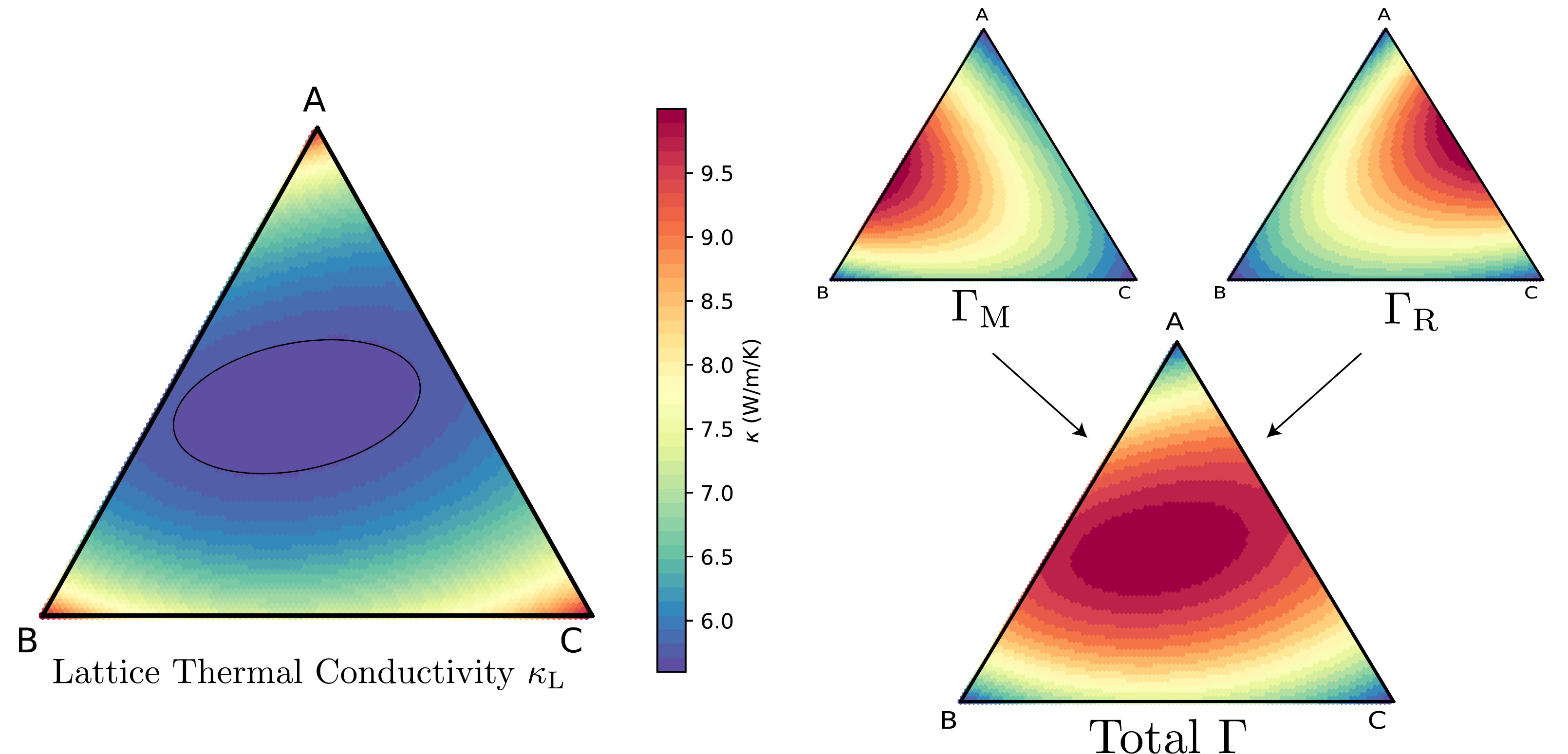}
    \caption{Example ternary alloy in which the mass contrast ($\Gamma\sub{M}$) and strain contrast ($\Gamma\sub{R}$) are maximized along different binary systems. The total scattering parameter $\Gamma$ is then peaked in the middle of the ternary alloy space, such that the minimum thermal conductivity region (encircled) is also centered around the equiatomic composition.}
    \label{fig:toy_alloy}
\end{figure}

\section{Origin of Sublattice Separation in Phonon Scattering}\label{supp:tamura}

Multi-sublattice substitution is discussed in the main text as one multicomponent alloying strategy for the reduction of thermal conductivity. The benefits of the approach are demonstrated in the \ch{(Pb,Sn)(Te,Se)} systemwhere the cation and anion site are both alloyed, and the minimum thermal conductivity occurs at the equiatomic \ch{Pb_{0.5}Sn_{0.5}Te_{0.5}Se_{0.5}} composition. This strategy suggests that the scattering along different sublattices should be uncorrelated, and confirms the importance of sublattice-specific models for phonon--point-defect scattering. The original proposed picture by Klemens of a monatomic lattice, where the atoms in the primitive unit cell are treated as a single, large vibrating mass, is then notably misleading. The point-defect scattering expression of Tamura (shown below) provides insight into why individual sublattices are decoupled from standpoint of phonon scattering. Here, $s = 1, 2,... N$ indexes the atom sites in the primitive unit cell while $i$ indexes the the atomic species that can occupy that site, including the host atom and any impurity atoms.

\begin{equation}
\Gamma^\mathrm{T}\sub{M} = \sum_s \sum_i f_{i,s}(\frac{M_{i,s} - \overline{M_s}}{\overline{M_s}})^2|(\mathbf{e_q}(s)\cdot \mathbf{e_{q'}}(s))|^2\nonumber
\end{equation}

The mass contrast term is weighted by the dot product of the polarization eigenvectors of site $s$ as it participates in the incident and final phonon mode. The monatomic lattice expression, in contrast, has no polarization vector dependence. This weighting by eigenvector overlap (related to vibrational amplitudes) causes there to be a frequency window for each sublattice in which point defect scattering is most effective. It is these unique frequency windows that can account for the decoupling (although not complete orthogonality) of the sublattices. The eigenvector overlap factor approaches the squared sublattice mass ($\overline{M_s}^2$) at the low-frequency limit. Therefore, compounds with low mass contrast between sublattices will behave more like monatomic lattices and the orthogonality of sublattices will likely diminish. To illustrate this decoupling, the phonon bandstructure of \ch{NaCl} is shown in colored by the eigenvector overlap factor $\phi_{\mathbf{q}}(s) = \frac{1}{N}\sum_{\mathbf{q'}}|\mathbf{e_q}(s)\cdot \mathbf{e_{q'}}(s)|^2$, for $s$ = the \ch{Na} site. The plot for $s$ = the \ch{Cl} site is exactly the opposite heatmap $\phi_{\mathbf{q}}(s = \mathrm{Cl}) = 1 - \phi_{\mathbf{q}}(s = \mathrm{Na})$. As shown, the frequency windows for scattering on the sublattice (indicated by a larger $\phi_{\mathbf{q}}(s)$ weighting) are separated such that the lighter element tends to dominate the higher frequency range.

\begin{figure}
    \centering
    \includegraphics[width = 0.7\textwidth]{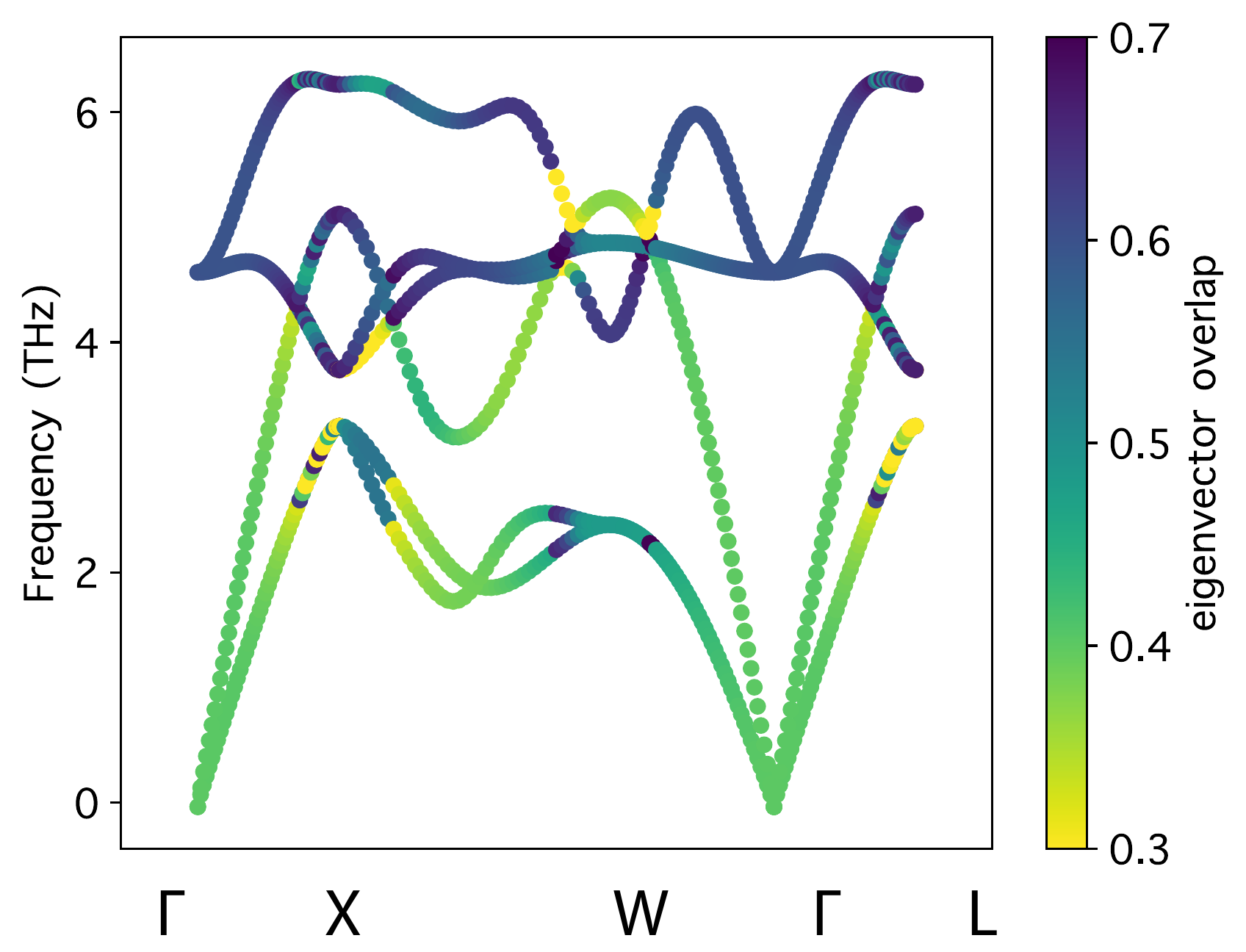}
    \caption{\textbf{Example Phonon Bandstructure Colored by Tamura Model Weighting Term} The phonon bandstructure of NaCl, in which phonon states are colored by the eigenvector overlap factor $\phi_{\mathbf{q}}(s)$, which weights the mass difference scattering term in the Tamura model. In this case, the bandstructure is colored by the $\phi_{\mathbf{q}}(s = \mathrm{Na})$ term, and as the lighter element, \ch{Na} follows the expected behavior by showing higher participation in the higher frequency range. The $s=\mathrm{Cl}$ case is the exact negative of the heatmap shown, such that the scattering frequency window is in the lower frequency range.}
    \label{fig:NaCl_overlap}
\end{figure}

\bibliographystyle{unsrt}
\bibliography{TECCA}
\end{document}